\documentclass[aps,prb,amsmath,amssymb,longbibliography,twocolumn,10pt,superscriptaddress]{revtex4-2}
\usepackage{graphicx}
\usepackage{bm}
\usepackage{xcolor}
\usepackage{mathrsfs}
\usepackage{amsthm}

\usepackage{amsmath}
\usepackage{overpic}
\usepackage[colorlinks,bookmarks=true,citecolor=blue,linkcolor=red,urlcolor=blue]{hyperref}
\usepackage[capitalise]{cleveref}

\crefname{appendix}{Appendix}{Appendices}
\Crefname{appendix}{appendix}{appendices}

\usepackage{physics}
\usepackage{mathtools}
\usepackage{orcidlink}
\usepackage{booktabs}

\newcommand{\up}{\ensuremath{\uparrow}}
\newcommand{\dn}{\ensuremath{\downarrow}}
\newcommand{\rh}[2]{\ensuremath{\bar \rho^{#1}_{#2}}}

\newcommand{\swedgesq}[2]{\ensuremath{\sin^2\left(\frac{#1 \wedge #2}{2}\right)}}

\makeatletter
\newcommand{\ssymbol}[1]{^{\@fnsymbol{#1}}}
\makeatother

\graphicspath{{./figures/}}
\begin{document}

\title{Sum rules and density-wave modes in spin-singlet fractional quantum Hall fluids}
 
\author{Ritajit Kundu\orcidlink{0000-0003-4917-4020}}
\email{ritajitk@imsc.res.in}
\affiliation{Institute of Mathematical Sciences, CIT Campus, Chennai, 600113, India}

\author{Rakesh K. Dora\orcidlink{0009-0009-0043-2982}}
\email{rakeshdora@iitb.ac.in}
\affiliation{Department of Physics, Indian Institute of Technology Bombay, Mumbai, 400076, India}

\author{Dung Xuan Nguyen\orcidlink{0000-0002-8595-0528}}
\email{dungmuop@gmail.com}
\affiliation{Institute for Interdisciplinary Research in Science and Education, ICISE, Quy Nhon, Vietnam}

\author{Ajit C. Balram\orcidlink{0000-0002-8087-6015}}
\email{cb.ajit@gmail.com}
\affiliation{Institute of Mathematical Sciences, CIT Campus, Chennai, 600113, India}
\affiliation{Homi Bhabha National Institute, Training School Complex, Anushaktinagar, Mumbai 400094, India} 

\begin{abstract}
Fractional quantum Hall (FQH) states are prototypical examples of strongly interacting topologically ordered systems. In this work, we obtain thermodynamic fits on the plane for the pair correlation function, and its Fourier transform, the static structure factor, of two-component spin-singlet Halperin and Jain FQH fluids by expanding them in the recently introduced basis of the orthogonal associated Laguerre polynomials [\href{https://doi.org/10.21468/SciPostPhys.14.6.149}
{Fulsebakke \emph{et al.}, SciPost Phys. \textbf{14}, 149 (2023)}] and ascertaining the expansion coefficients by fitting them to large-system Monte Carlo data evaluated using their trial wavefunctions. In this fitting procedure, aside from constraining the exact short-distance behavior of the wavefunction, we also derive and enforce the sum rules that the long-wavelength expansion of the static structure factor must adhere to. We show that incorporating these constraints is crucial for obtaining numerically stable and accurate values of the long-wavelength Girvin-MacDonald-Platzman (GMP)/symmetric density-wave excitation gap. We further extend this approach to spin-resolved density-correlation functions, enabling the evaluation of the gap of the antisymmetric density-wave mode for these spin-singlet FQH states. Finally, we use the density-correlators to compute variational energies of the states and construct phase diagrams for bilayer FQH systems. These could be relevant for understanding recent bilayer FQH experiments that map out the phase diagram by tuning the interlayer separation and density-imbalance/layer-polarization.
\end{abstract}

\maketitle

\section{Introduction}
The fractional quantum Hall (FQH) effect is a paradigmatic example of a strongly correlated quantum phase of matter realized in a two-dimensional electron gas subjected to a strong perpendicular magnetic field cooled to cryogenic temperatures~\cite{Tsui82, Laughlin83}. FQH states are realized at a partial filling of a Landau level (LL), wherein Coulomb interactions conspire to open an incompressibility gap at certain rational fillings of the LL. The FQH states are topologically ordered phases that support quasiparticle excitations with fractional charge and anyonic braiding statistics~\cite{Wen1990, Wen92a, Wen92b, Wen95, Nayak08, Nakamura20, Bartolomei20, Greiter24}. Although most FQH states are fully spin-polarized due to the large external magnetic field, several experimentally observed FQH states are known to be spin-singlets or partially spin-polarized~\cite{Eisenstein90, Kang97, Chang03b, Tracy07, Pan12, Feldman13, Liu15, Huang21, Huang25}. The additional spin degree of freedom gives rise to richer correlation effects and supports a wider variety of collective excitations, making spinful FQH states a versatile platform for exploring multicomponent topological phases~\cite{Halperin83, Rasolt86, Yoshioka88, Yoshioka89, Park98, Girvin07, Dean11, Zaletel15, Balram15, Balram15a, Liu15}.

Neutral collective excitations provide important information about the internal dynamics of quantum Hall fluids and serve as sensitive probes of their underlying topological order~\cite{Kallin84, Girvin85, Girvin86, Rasolt86, Liu20, Balram21d, Nguyen22, Liang24, Bose25, Dora25}. Girvin, MacDonald, and Platzman (GMP)~\cite{Girvin85, Girvin86} used a single-mode approximation (SMA) to construct a density-wave ansatz which accurately captures the lowest-lying branch of neutral excitations, particularly in the long-wavelength limit, in many FQH fluids~\cite{Girvin85, Girvin86, Balram24, Dora24, Dora25}. When the spin degree of freedom is active, FQH states can support a multitude of collective excitations. Among the simplest density-wave modes are the symmetric density-wave (SDW), which is equivalent to the GMP mode~\cite{Girvin85, Girvin86}, and antisymmetric density-wave (ASDW) modes~\cite{Rasolt86, Dora25}, that correspond to the in-phase and out-of-phase density fluctuations of the two spin components, respectively. These excitations can be directly probed using inelastic Raman scattering or surface-acoustic wave experiments~\cite{Pinczuk93, Song94, Kukushkin09, Nguyen21a, Liang24, Yang26}, providing a useful tool for distinguishing competing candidate states at a given filling.

Accurate calculations of long-wavelength collective excitation gaps remain challenging. Exact diagonalization provides the full excitation spectrum but is limited to relatively small system sizes~\cite{He94, Morf02, Balram13}. Density-matrix renormalization group calculations can accurately determine ground states for much larger systems, but extracting neutral collective-mode dispersions is considerably more difficult, particularly in multicomponent systems~\cite{Zhao11, Johri14, Zaletel15, Kumar24a}. These limitations motivate approaches that determine the excitation spectrum directly from accurate trial wavefunctions. A nice feature of the aforementioned density modes is that their gaps can just be obtained from an evaluation of the density-density correlation function, i.e., the pair-correlation function $g(r)$, or its Fourier transform, the static structure factor, $S(q)$, in the ground state~\cite{Girvin85, Girvin86, Feynman98}. 

Girvin~\cite{Girvin84a} introduced a set of basis functions to model the $g(r)$ of an incompressible FQH state. The expansion coefficients have to adhere to certain constraints imposed by the short-distance vanishing properties of the specific trial wavefunction under consideration, and the exact sum rules that follow from the incompressibility and topological properties of the FQH state that determine the leading long-wavelength expansion of $S(q)$~\cite{Stillinger68a, Stillinger68b, Girvin85, Kalinay00, Gail08, Nguyen17}. Imposing these constraints and fitting against the Monte Carlo data of the trial wavefunction determines the expansion coefficients. However, Girvin's expansion employs a non-orthogonal basis, leading to numerical instabilities in the fitting procedure. More recently, Fulsebakke \textit{et al.}~\cite{Fulsebakke23} introduced an orthogonal basis that significantly improves numerical stability. While this method accurately reproduced the $g(r)$ and the roton minimum in the density-wave excitation, it did not enforce the complete set of constraints that the long-wavelength expansion of $S(q)$ must respect. Consequently, the correct long-wavelength density-wave excitation gap was not reproducible by this procedure.

In this work, we improve the fitting procedure by using the orthogonal basis, while augmenting it with necessary constraints to ensure the correct leading long-wavelength expansion of $S(q)$. We derive the relevant constraint equations that the fitting parameters must satisfy and enforce them during the fitting procedure. This guarantees that the fitted $g(r)$ and $S(q)$ satisfy both the exact short-distance behavior and all the known long-wavelength sum rules. The resulting parameterization is numerically stable and enables a reliable calculation of the long-wavelength density-wave excitation gaps. As we show below, although the sum-rule constraints have only a minor effect on the ground-state energies of candidate states, they are imperative for obtaining the correct long-wavelength density-wave excitation gaps.

We further derive the exact sum rules for the spin-resolved static structure factors for spinful FQH states using effective field theory~\cite{Wen92a, Wen95, Lopez01, Nguyen17} and show that they agree with previous results obtained from the plasma analogy~\cite{Gail08}. Motivated by these results, we extend the orthogonal-basis expansion framework to spin-resolved pair-correlation functions by incorporating the corresponding long-wavelength constraints. This provides a systematic and accurate method for computing the SDW and ASDW excitation gaps of spin-singlet FQH states~\cite{Rasolt86, Dora25}. We apply our framework to the prominent Halperin~\cite{Halperin83} and Jain~\cite{Jain89} spin-singlet FQH states at fillings $\nu_{b}=2/3$ for bosons and $\nu=2/3$ and $2/5$ for fermions. An important caveat to keep in mind is that the density-wave modes are not always the lowest-lying neutral excitations~\cite{Balram21d, Balram24, Dora25}, and in such scenarios, one has to resort to other means to construct the lowest-lying neutral excitations.
 
Additionally, we use the $g(r)$ and $S(q)$ we obtained to compute the variational energies of these singlet states in a bilayer setting, where the layer index acts as a pseudospin degree of freedom~\cite{Halperin83, Moon95, Shi08, Eisenstein14, Geraedts15, Peterson15, Faugno20}. By comparing the energies of competing candidate states as functions of the interlayer separation and layer-polarization/density-imbalance, we construct variational phase diagrams that identify the phases expected to be stabilized at a given filling. These results are directly relevant to the recent bilayer FQH experiments that tune the interlayer separation and density imbalance to study the competition between different phases~\cite{Wagner2024, Nguyen24a, Nguyen26}.

The remainder of this paper is organized as follows. In \cref{sec:model}, we provide the necessary background by introducing the lowest LL (LLL) projected density operators, their commutator algebra, the LLL-projected Hamiltonian, and the trial wavefunctions for the spin-singlet FQH states considered here. In \cref{sec:pair}, we review the density-density correlation functions and the sum rules that they respect. In \cref{sub:field-theory}, we derive the sum rules for spin-resolved density-density correlators for spinful FQH states using the effective Chern-Simons-Maxwell theory. In \cref{sec:collective}, we obtain the gap equation for the SDW and ASDW modes for spinful FQH states and the GMP and spin-flip modes for fully polarized states. In \cref{sec:fitting}, we review Girvin's expansion, introduce the orthogonal basis of Fulsebakke \textit{et al.}, and derive the constraint equations used in the fitting procedure. In \cref{sec:results-pair-corr-SF}, we present the fitted density-density correlation functions, and in \cref{sec:results-collective-excitations}, we present the dispersions of the collective density-wave modes obtained from them. In \cref{sec:phase-diagram}, we compare the energies of spin-polarized, spin-singlet, and layer-decoupled candidate states to construct variational phase diagrams at certain fillings. Finally, \cref{sec:conclusion} summarizes our main results and discusses future directions.

\section{Model and formulation}
\label{sec:model}
In this section, we introduce the spin-resolved density operators and their LLL-projected forms, along with their commutation relations. We then give the LLL-projected interaction Hamiltonian and review the trial wavefunctions considered in this work.

\subsection{Spinful density operators on the sphere}
We consider a system of $N$ spin-$1/2$ particles confined to a two-dimensional plane. Following Refs.~\cite{Rasolt86, Dora25}, we define the spin-resolved density operators as
\begin{align}
    \rho^\alpha(r) =  \sum_{i=1}^{N}
    \sigma_i^\alpha
    \otimes \delta^{(2)}(\bm r - \bm r_i),
\end{align}
where $\sigma_i^\alpha$ for $\alpha{=}I,x,y,z$ are Pauli matrices ($\sigma^I$ is the $2{\times}2$ identity matrix) that act on the spin of the $i^{\rm th}$ particle. The position of the $i^{\rm th}$ particle is given by $\bm r_i = (\mathsf x_i, \mathsf y_i)$.
The corresponding Fourier-space spin-resolved density operators are
\begin{align}
    \rho_q^\alpha
    =
    \sum_{i=1}^{N}
    \sigma_i^\alpha
    \otimes
    \exp\!\left[
        \frac{\iota}{2}
           \bm q \cdot \bm r_i
    \right],
\end{align}
where $\iota=\sqrt{-1}$ is the imaginary unit. Throughout this work, we employ the complex representation for the coordinates and momenta, i.e., for the $i^{\rm th}$ particle we use coordinates $z_i = \mathsf x_i - \iota \mathsf y_i$, and for momentum we use $q{=}q_x{+}\iota q_y$. In terms of these, the density operator is
\begin{align}
    \rho_q^\alpha
    =
    \sum_{i=1}^{N}
    \sigma_i^\alpha
    \otimes
    \exp\!\left[
        \frac{\iota}{2}
        \left(
            q z_i^*
            +
            q^* z_i
        \right)
    \right],
    \label{eq:density-operator}
\end{align}
where the superscript $^{*}$ denotes complex conjugation. The LLL-projected density operator, $\bar{\rho}_q^\alpha$, is obtained using the Girvin-Jach projection scheme~\cite{Girvin84b}, wherein all the $z^{*}$ are moved all the way to the left, and then replaced by $2\partial/\partial z$, i.e., the substitution $z^*{\rightarrow}2\partial/\partial z{\equiv}2\partial_{z}$ is done after bringing all the $z^{*}$ to the extreme left. When acting on a LL wavefunction, the resulting derivatives act only on its polynomial part and not on the ubiquitous Gaussian factor (which we will omit for ease of notation). The projection ensures that the action of the projected density operator on a LLL state yields another state that remains entirely within the LLL. Consequently,
\begin{align}
    \bar{\rho}_q^\alpha
    &=
    \sum_{i=1}^{N}
    \sigma_i^\alpha
    \otimes
    \mathcal{P}_{\mathrm{LLL}}
    \exp\!\left[
        \frac{\iota}{2}
        \left(
            q z_i^* + q^* z_i
        \right)
    \right]
    \mathcal{P}_{\mathrm{LLL}}
    \\
    &=
    e^{-q^2/4}
    \sum_{i=1}^{N}
    \sigma_i^\alpha
    \otimes
    \tau_q(i),
    \label{eq:projected-density-operator}
\end{align}
where $\mathcal{P}_{\mathrm{LLL}}$ is the projection operator onto the LLL and
$
    \tau_q(i)
    {=}
    \exp\!\left(
        {-}\iota q \partial z_i
        {-} \iota q^* z_i/2
    \right)
$
is the magnetic translation operator acting on the $i^{\rm th}$ particle. These operators satisfy the algebra
\begin{align}
    [\tau_q,\tau_k]
    =
    2 \iota\,
    \tau_{q+k}
    \sin\!\left(
        \frac{q \wedge k}{2}
    \right),
\end{align}
where the wedge product 
$
q \wedge k
    =
    \ell^2
    (\bm q \times \bm k)
    \cdot
    \hat{\bm z}
$
with $\ell=\sqrt{\hbar c/(eB)}$ being the magnetic length at the magnetic field $B$. The projected spin-resolved density operators obey the following closed algebra
\begin{align}
    [\bar{\rho}_q^\alpha,\bar{\rho}_k^\beta]
    &=
    2
    e^{\bm q\cdot\bm k/2}
    \Bigg[
    \iota
    \sin\!\left(
        \frac{q \wedge k}{2}
    \right) \times
    \nonumber\\
    \Big(
        \delta_{\alpha\beta}
        \bar{\rho}_{q+k}^{I}
    &
        +\,
        \delta_{\alpha I}
        (1-\delta_{\beta I})
        \bar{\rho}_{q+k}^{\beta}
        +
        (1-\delta_{\alpha I})
        \delta_{\beta I}
        \bar{\rho}_{q+k}^{\alpha}
    \Big)
    \nonumber\\
    &\qquad
    +
    \cos\!\left(
        \frac{q \wedge k}{2}
    \right)
    \sum_{\gamma}
    \epsilon_{I\alpha\beta\gamma}
    \bar{\rho}_{q+k}^{\gamma}
    \Bigg].
    \label{eq:spinful-commutator-algebra}
\end{align}
Here, $\varepsilon_{I\alpha\beta\gamma}$ is the fully anti-symmetric Levi-Civita tensor with $\varepsilon_{Ixyz}{=}1$. For $\alpha{=}\beta{=}I$, the expression in Eq.~\eqref{eq:spinful-commutator-algebra} reduces to the GMP algebra~\cite{Girvin85, Girvin86}. The spin-up and spin-down projected density operators are
\begin{align}
\bar\rho_k^\uparrow
&=
e^{-k^2/4}
\sum_i
\left(\frac{\sigma_i^I+\sigma_i^z}{2}\right)
\otimes
\tau_k(i),
\\
\bar\rho_k^\downarrow
&=
e^{-k^2/4}
\sum_i
\left(\frac{\sigma_i^I-\sigma_i^z}{2}\right)
\otimes
\tau_k(i),
\end{align}
and one can check that $\bar\rho_k^I= \bar\rho_k^\uparrow+\bar\rho_k^\downarrow$ and $\bar\rho_k^z= \bar\rho_k^\uparrow-\bar\rho_k^\downarrow$.

\subsection{Lowest-Landau-level projected Hamiltonian}

Throughout this work, we consider particles confined to the LLL. In this setting, the kinetic energy is quenched and contributes only an overall constant that we set as the baseline to measure energies relative to. Thus, the Hamiltonian consists solely of the interaction term,
\begin{align}
    H = \frac12 \sum_{i\neq j} v(\abs{\bm r_i-\bm r_j}),
\end{align}
where $v(r)$ is chosen to be a central potential. In terms of the total density operator $\rho^I(\bm r)$, a spin-independent $SU(2)$-symmetric Hamiltonian reads
\begin{align}
    H = \frac12 \int \dd\bm r\, \dd\bm r' \, v(\bm r-\bm r') : \rho^I(\bm r) \rho^I(\bm r') : ,
\end{align}
where the normal ordering $:~:$ removes self-interaction terms. In Fourier space,
\begin{align}
    H = \frac12 \int \frac{\dd\bm q}{(2\pi)^2} \, v(q) : \rho^I_{-q} \rho^I_q :,
\end{align}
where $v(q)$ is the Fourier transform of $v(r)$. Using $: \rho^I_{-q} \rho^I_q : {=} \rho^I_{-q} \rho^I_q - N,$ the Hamiltonian becomes
\begin{align}
    H = \frac12 \int \frac{\dd\bm q}{(2\pi)^2} \, v(q) \left( \rho^I_{-q} \rho^I_q - N \right).
\end{align}

Projecting onto the LLL amounts to replacing the density operators by their projected versions, which results in~\cite{Girvin86} (see Appendix~\ref{app:Relation between the projected and unprojected structure factors}),
\begin{align}
    \bar H
    =
    \frac12
    \int
    \frac{\dd\bm q}{(2\pi)^2}
    \,
    v(q)
    \left(
        \bar\rho^I_{-q}
        \bar\rho^I_q
        -
        N e^{-q^2/2}
    \right).
    \label{eq:projected-Hamiltonian}
\end{align}

We will consider both the Coulomb interaction,
\begin{align}
    v^{\rm C}(r)=\frac{1}{ r},
    \qquad
    v^{\rm C}(q)=\frac{2\pi}{ q},
\end{align}
and the contact interaction~\cite{Trugman85},
\begin{align}
    v^{\rm CI}(r)
    =
    4\pi
    \delta^{(2)}(\bm r),
    \qquad
    v^{\rm CI}(q)
    =
    4\pi.
\end{align}

A rotationally invariant interaction projected to the LLL is conveniently characterized in terms of its Haldane pseudopotentials $\{V_{\mathfrak m}\}$, which are the energies of two particles with relative angular momentum $\mathfrak m$~\cite{Haldane83}. For an interaction $v(q)$,
\begin{align}
    V_{\mathfrak m} = \int \frac{\dd\bm q}{(2\pi)^2} \, v(q) e^{-q^2} L_{\mathfrak m}(q^2),
\end{align}
where $L_{\mathfrak m}(x)$ is the Laguerre polynomial of degree $\mathfrak m$. Conversely, an arbitrary rotationally invariant interaction can be reconstructed from its pseudopotentials as
\begin{align}
    v(r) &= 4\pi\ell^2 \sum_{\mathfrak m=0}^{\infty} V_{\mathfrak m} L_{\mathfrak m}(-\ell^2\nabla^2) \delta^{(2)}(\bm r),\\
    v(q) &= 4\pi\ell^2 \sum_{\mathfrak m=0}^{\infty} V_{\mathfrak m} L_{\mathfrak m}(\ell^2q^2).
\end{align}
The pseudopotentials of the Coulomb interaction are
$
V_{\mathfrak m}^{\rm C}
{=}\Gamma(\mathfrak m+1/2)/(2\,\mathfrak m!)
$, 
whereas the contact interaction corresponds to a $\mathfrak{m}$-only pseudopotential, i.e., 
$
V_{\mathfrak m}^{\mathrm{CI}}=V_0\delta_{\mathfrak m,0},
$
where $V_0$ quantifies its strength.  

\subsection{Candidate ground-state wavefunctions for spin-singlet fractional
quantum Hall states}

In this work, we consider spin-singlet FQH ground states with total spin quantum numbers $\mathbb{S}{=}\mathbb{S}^z{=}0$. We model these states via the Halperin~\cite{Halperin83} and Jain~\cite{Jain89} trial wavefunctions, which we describe next. 

The Halperin-$(m,m,n)$ state~\cite{Halperin83} at filling factor $\nu=2/(m+n)$ is described by the wavefunction
\begin{align}
\label{eq: Halperin_mmpn_wavefunction}
    \Psi_{\nu=2/(m+n)}^{H-(m,m,n)}
    =
    \left(\Phi^\uparrow_{\mathrm{intra}}\right)^m
    \left(\Phi^\downarrow_{\mathrm{intra}}\right)^m
    \left(\Phi_{\mathrm{inter}}\right)^n,
\end{align}
where
\begin{align}
    \Phi^\uparrow_{\mathrm{intra}}
    &= \prod_{1 \leq i < j \leq N_\uparrow}
    (z_i^\uparrow - z_j^\uparrow), \\
    \Phi^\downarrow_{\mathrm{intra}}
    &= \prod_{1 \leq i < j \leq N_\downarrow}
    (z_i^\downarrow - z_j^\downarrow), \\
    \Phi_{\mathrm{inter}}
    &= \prod_{i=1}^{N_\uparrow}
    \prod_{j=1}^{N_\downarrow}
    (z_i^\uparrow - z_j^\downarrow),
\end{align}
where $N_{\uparrow}$ and $N_{\downarrow}$ are the number of $\uparrow$ and $\downarrow$ particles, respectively. This state has Wen-Zee shift $\mathcal{S}=m$~\cite{Wen92} and is characterized by the $K$ matrix~\cite{Wen95}
\begin{align}
    K=
    \begin{pmatrix}
        m & n\\
        n & m
    \end{pmatrix},
    \label{eq:Halperin-Kmatrix}
\end{align}
with charge vector $\mathbf {t}=(1,1)^T$ and spin vector $\mathbf{s}=(m/2,m/2)^T$. Among these, the Halperin-$(m,m,m{-}1)$ states are spin-singlets~\cite{Yoshioka89}. In this work, we focus on the bosonic Halperin-$(2,2,1)$ state at filling $\nu_b=2/3$ and the fermionic Halperin-$(3,3,2)$ state at filling $\nu=2/5$.

We will also consider the fermionic Jain spin-singlet state~\cite{Jain89} at filling $\nu=2/3$, which has $\mathcal{S}=1$. Its wavefunction is given by
\begin{align}
\label{eq: wfn_Jain_2_3_spin_singlet}
    \Psi_{\nu=2/3}^{\mathrm{Jain~singlet}}
    =
    \mathcal{P}_{\mathrm{LLL}}
    \Phi_1^2
    \left[\Phi^\uparrow_{\mathrm{intra}}\Phi^\downarrow_{\mathrm{intra}}\right]^*,
\end{align}
where $\Phi_1=\Psi_{\nu=1}^{H-(1,1,1)}$ is the Laughlin-Jastrow factor or, equivalently, the wavefunction of the filled LLL. The projection to the LLL is carried out using the Jain-Kamilla method, details of which are given in Refs.~\cite{Jain97, Davenport12}. The Jain spin-singlet $2/3$ state lies in the same universality class as the Halperin-$(1,1,2)$ state. Still, unlike the Halperin-$(1,1,2)$ state, which phase separates due to stronger inter-correlations compared to the intra-ones~\cite{Gail08, Simon25}, the wavefunction of Eq.~\eqref{eq: wfn_Jain_2_3_spin_singlet} describes a uniform FQH liquid at $2/3$. We note that the fermionic Halperin-$(3,3,2)$ state at filling $\nu=2/5$ is identical to the fermionic Jain spin-singlet state at $2/5$, the wavefunction for which is obtained by removing the complex conjugation in Eq.~\eqref{eq: wfn_Jain_2_3_spin_singlet}.

Throughout the remainder of the paper, quantities associated with a particular trial wavefunction are denoted by a superscript labeling the state, and a subscript specifying the filling. Next, we define the pair-correlation functions and static structure factors that are computed numerically from these states and subsequently used to determine the density-wave dispersions and ground state energies. 

\section{Pair correlation and structure factor}
\label{sec:pair}
In this section, we define the spin-resolved pair-correlation function and its Fourier transform, the static structure factor. We then use the plasma analogy and effective field theory to constrain the leading long-wavelength behavior of the structure factors.

\subsection{Pair correlation function}
For a translationally and rotationally invariant normalized \(N\)-particle state \(|\Psi\rangle\) at filling $\nu$ and density \(\rho_0{=}\nu / 2\pi\) (we set $\ell{=}1$), the $g(r)$ is obtained by integrating the probability density \(|\Psi|^2\) over the coordinates of all particles except two, which are fixed at positions \(\bm r\) and \(\bm 0\),
\begin{align}
\label{eq:define_pair_correlation}
g\left(r\right)&=\frac{N(N-1)}{\rho_0^2}\int \dd^{2}
\bm{r}_{3}{\cdots}\dd^{2}\bm{r}_{N} \sum_{\sigma_{1},{\cdots},\sigma_{N}} {\times}  \\
&|\Psi(\{\bm{r},\sigma_1\}, \{\bm{0},\sigma_2\}, \{\bm{r}_{3},\sigma_{3}\},{\cdots},\{\bm{r}_{N},\sigma_{N}\} )|^{2}.\nonumber
\end{align}
The $g(r)$ encodes the probability of finding a particle at a distance $r$ away from a given particle that is chosen to lie at the origin. 

The spin-resolved $g^{\alpha, \beta}(r)$ is defined similarly, with the two fixed particles constrained to have spins \(\alpha\) and \(\beta\), as
\begin{align}
\label{eq:define_pair_correlation_spin}
g^{\alpha, \beta}\left(r\right)&=\frac{N(N-1)}{\rho_{0,\alpha} \rho_{0, \beta}}\int \dd^{2}
\bm{r}_{3}{\cdots}\dd^{2}\bm{r}_{N} \sum_{\sigma_{3},{\cdots},\sigma_{N}} {\times}  \\
&|\Psi(\{\bm{r},\alpha\}, \{\bm{0},\beta\}, \{\bm{r}_{3},\sigma_{3}\},{\cdots},\{\bm{r}_{N},\sigma_{N}\} )|^{2}.\nonumber
\end{align}

The normalization is such that \(g(r){\to}1\) and \(g^{\alpha,\beta}(r){\to}1\) as \(r{\to}\infty\) for all \(\alpha,\beta{\in}[\up,\dn]\). The $g(r)$ is related to $g^{\alpha, \beta}\left(r\right)$ as
\begin{align}
\label{eq: gr_in_terms_of_its_components}
    g(r)=\sum_{\alpha,\beta}\frac{\rho_{0,\alpha}\rho_{0,\beta}}{\rho_0^2}\,g^{\alpha,\beta}(r),
\end{align}
where \(\rho_{0,\alpha}\) is the density of particles with spin \(\alpha\).
The per-particle energy can be computed from $g^{\alpha, \beta}\left(r\right)$ for an arbitrary two-body interaction $v^{\alpha,\beta}(r)$ as~\cite{Giudici08, Kundu26}
\begin{align}
    E = \frac{\nu}{2} \sum_{\alpha,\beta}
\frac{\rho_{0,\alpha}\rho_{0,\beta}}{\rho_0^2}
    \int \dd r\, r\, v^{\alpha,\beta}(r)\,\left[g^{\alpha,\beta}(r)-1\right].
    \label{eq:energy-gr}
\end{align}
In Sec.~\ref{sec:phase-diagram}, we use this relation to evaluate the energies of candidate bilayer FQH states.

For spin-independent interactions, \(v^{\alpha,\beta}(r)=v(r)\); thus, we get the familiar result involving only \(g(r)\), i.e., 
\begin{align}
    E = \frac{\nu}{2} 
    \int \dd r\, r\, v(r)\,\left[g(r)-1\right].
    \label{eq:energy-gr-1}
\end{align}
Throughout, we quote energies in Coulomb units of $e^2/(\varepsilon\ell)$, where $\varepsilon$ is the dielectric constant of the background host material.

\subsection{Static structure factor}
The static structure factor provides the momentum-space description of density-density correlations and is obtained from the Fourier transform of the pair-correlation function as
\begin{subequations}
\begin{align}
    S(q)
    &=1+\rho_0\int \dd^2\bm r\,e^{\iota\bm q\cdot\bm r}[g(\bm r)-1], \label{eq:SF_from_pair}\\
    &=1+\rho_0\sum_{n=0}^{\infty}
    \frac{(-1)^n}{(n!)^2}
    \left(\frac{q}{2}\right)^{2n}
    \mathcal I_{2n},
    \label{eq:SF1}
\end{align}
\end{subequations}
where \(\mathcal I_{2n}=\int \dd^2\bm r\,r^{2n}h(r)\) is the $2n$-th moment of the total correlation function, \(h(r)=g(r)-1\), which measures the excess probability density of finding a particle at a distance $r$ away from a particle at the origin relative to the mean density. Similarly, the spin-resolved static structure factor, $S^{\alpha,\beta}(q)$, is obtained by Fourier transforming $g^{\alpha,\beta}(r)$ as
\begin{subequations}
\begin{align}
    S^{\alpha,\beta}(q)
    &=\frac{\rho_{0,\alpha}}{\rho_0}\delta_{\alpha,\beta}
    +\frac{\rho_{0,\alpha}\rho_{0,\beta}}{\rho_0}
    \int \dd^2\bm r\,e^{-\iota\bm q\cdot\bm r}h^{\alpha,\beta}(r), \label{eq:structure-factor}\\
    &=\frac{\rho_{0,\alpha}}{\rho_0}\delta_{\alpha,\beta}
    +\frac{\rho_{0,\alpha}\rho_{0,\beta}}{\rho_0}
    \sum_{n=0}^{\infty}
    \frac{(-1)^n}{(n!)^2}
    \left(\frac{q}{2}\right)^{2n}
    \mathcal I_{2n}^{\alpha,\beta},
    \label{eq:structure-factor1}
\end{align}
\end{subequations}
where \(\mathcal I_{2n}^{\alpha, \beta}=\int \dd^2\bm r\,r^{2n}h^{\alpha, \beta}(r)\) and \(h^{\alpha,\beta}(r)=g^{\alpha,\beta}(r)-1\). Analogous to Eq.~\eqref{eq: gr_in_terms_of_its_components} for $g(r)$, we have
\begin{align}
\label{eq: Sq_in_terms_of_its_components}
    S(q) = \sum_{\alpha,\beta}S^{\alpha,\beta}(q)
\end{align}
Next, we discuss the leading long-wavelength behavior of $S^{\alpha,\beta}(q)$, which provides constraints that need to be imposed while parameterizing $g^{\alpha,\beta}(r)$.

\subsection{Sum rules for the structure factor \texorpdfstring{$\bm{S(q)}$}{S(q)}}
\label{ssec: sum_rules_Sq}

The long-wavelength behavior of the $S(q)$ of FQH liquids is highly constrained since it needs to respect exact sum rules that the moments of its $g(r)$ need to adhere to. From \cref{eq:SF1}, $S(q)$ admits the $q{\to}0$ expansion
\begin{align}
    S(q)
    = s_0 + s_2 q^2 + s_4 q^4 + s_6 q^6 + \mathcal{O}(q^8),
    \label{eq:sexpansion}
\end{align}
with coefficients
\begin{subequations}
\begin{align} \label{eq:s_coeffs0}
    s_0 &= 1+\rho_0\mathcal I_0, \\
    s_2 &= -\frac{\rho_0\mathcal I_2}{4}, \\
    s_4 &= \frac{\rho_0\mathcal I_4}{64}, \\
    s_6 &= -\frac{\rho_0\mathcal I_6}{2304}.
    \label{eq:s_coeffs}
\end{align}
\end{subequations}

For FQH states that admit a plasma analogy~\cite{Laughlin83}, like the Halperin states of Eq.~\eqref{eq: Halperin_mmpn_wavefunction}, the long-range nature of the Coulomb interaction leads to a hierarchy of exact sum rules for the moments $\mathcal I_{2n}$~\cite{Kalinay00}. These sum rules determine the coefficients \(s_0,s_2,s_4,s_6\) via Eq.~\eqref{eq:s_coeffs0}-\eqref{eq:s_coeffs}.

We briefly review the plasma analogy for the $\nu=1/m$ Laughlin state~\cite{Laughlin83} that is described by the wavefunction
\begin{equation}
    \Psi^{\rm Laughlin}_{\nu=1/m}=\prod_{i<j}(z_i-z_j)^m.
\end{equation}
In the plasma analogy, the $|\Psi^{\rm Laughlin}_{\nu=1/m}|^2$, is interpreted as the Boltzmann weight $e^{-\beta H}$ of a two-dimensional one-component plasma~\cite{Laughlin83}, where
\begin{align}
    \beta H
    = -2m \sum_{i<j}\log|z_i-z_j|
    + \sum_i \frac{|z_i|^2}{2}.
\end{align}
The Coulomb interaction in the two-dimensional plasma is given by $\beta v(z_i,z_j)=-\Gamma\log|z_i-z_j|$, with coupling constant $\Gamma=2m$ for the Laughlin state.

The total correlation $h(r)$ is related to the direct correlator $c(r)$ via the Ornstein-Zernike (OZ) equation
\begin{align}
    h(r)
    = c(r)
    + \rho_0\int \dd^2\bm r'\,
    c(|\bm r-\bm r'|)\,
    h(r').
\end{align}
Taking its Fourier transform gives
\begin{align}
    h(q)
    &= \frac{c(q)}{1-\rho_0c(q)},\\
    \shortintertext{from which $S(q)$ is obtained as}
    S(q)
    &= 1+\rho_0h(q)
    = \frac{1}{1-\rho_0c(q)}.
    \label{eq:SF-OZ}
\end{align}

Using diagrammatic methods, Kalinay \emph{et al.}~\cite{Kalinay00} obtained the small-$q$ expansion of $c(q)$ up to order $q^2$ as
\begin{align}
    c(q)
    = -\frac{2\pi\Gamma}{q^2}
    + \frac{2\pi\Gamma}{8\pi\rho_0}
    - \frac{2\pi q^2}{96(\pi\rho_0)^2}
    + \mathcal{O}(q^4).
    \label{eq:direct-correlation}
\end{align}
The leading singular term is simply the Fourier transform of the Coulomb interaction, i.e., $-\beta v(q)=-\Gamma/q^2$, from which one immediately obtains $S(q)=q^2/2+\mathcal{O}(q^4)$. More generally, knowledge of the expansion of $c(q)$ through order $q^{2j}$ determines the small-$q$ expansion of $h(q)$, and hence $S(q)$, to order $q^{2(j+2)}$ via Eq.~\eqref{eq:SF-OZ}. Equivalently, it determines the moments up to $\mathcal I_{2(j+2)}$.

Substituting the expansion of \cref{eq:direct-correlation} into \cref{eq:SF-OZ} and matching like powers of $q$ yields
\begin{subequations}
\begin{align}
    s_0 &= 0,\\
    s_2 &= \frac{1}{2}, \label{eq:s2_value}\\
    s_4 &= \frac{\Gamma-4}{4(\Gamma\nu)^2},\\
    s_6 &= \frac{(\Gamma-6)(3\Gamma-8)}
    {48(\Gamma\nu)^3}.
\end{align}
\end{subequations}
The condition $s_0{=}0$ is a consequence of charge neutrality of the plasma, while $s_2{=}1/2$ follows from the Stillinger-Lovett sum rule~\cite{Stillinger68a, Stillinger68b} which expresses its perfect screening. The two coefficients $s_0$ and $s_2$ are universal and do not depend on the nature of the FQH state. The coefficient $s_4$ is fixed by the compressibility sum rule~\cite{Vieillefosse75, Baus78}, while the constraint on $s_6$, derived in Ref.~\cite{Kalinay00}, is commonly referred to as the sixth-moment sum rule, which does not lend itself to a simple physical interpretation. For the $\nu=1/m$ Laughlin state, where $\Gamma=2m$, $s_4=(m-2)/8$ and $s_6=(3m-4)(m-3)/96$.

Refs.~\cite{Can14, Can15, Nguyen17} extended the $q{\to}0$ expansion of $S(q)$ to more general LLL chiral FQH states that do not admit a plasma description using wavefunctions in curved geometry and inhomogeneous magnetic fields. This geometric-response theory found that the expansion coefficients are entirely determined from the topological quantum numbers of the FQH state, namely the filling fraction $\nu$, Wen-Zee shift $\mathcal{S}$~\cite{Wen92}, chiral central charge $c_-$~\cite{Kane97}, and orbital-spin variance $\mathrm{var}(s)$~\cite{Gromov15, Bradlyn15}. For Abelian chiral FQH states, these quantities can be expressed in terms of the $K$ matrix, charge vector $\mathbf {t}$, and spin vector $\mathbf{s}$~\cite{Wen95} as~\cite{Nguyen17}
\begin{subequations}
\label{eq:s4-s6-derive}
\begin{align}
    s_4 &= \frac{\mathcal S-2}{8},\\
    s_6 &= \frac{2-\mathcal S}{16}
    - \frac{b}{8\nu},
\end{align}
\end{subequations}
where
\begin{align}
    b
    = \nu\frac{\mathcal S}{2}
    \left(1-\frac{\mathcal S}{2}\right)
    + \frac{c_{-}}{12}
    -\nu {\rm var}(s).
\end{align}
Here $\mathcal S=2 \mathbf t^T K^{-1}\mathbf{s}/\nu$ is the Wen-Zee shift~\cite{Wen92}, $c_-$ is the chiral central charge~\cite{Kane97}, and $\mathrm{var}(s)=\mathbf s^T K^{-1}\mathbf s-\nu(\mathcal S/2)^2$ is the orbital-spin variance~\cite{Gromov15, Bradlyn15}. For the $\nu=1/m$ Laughlin state, $\mathcal S=m$, $c_{-}=1$, and ${\rm var}(s)=0$ recovering the same values of $s_4$ and $s_6$ as those obtained from the OZ analysis.

\subsection{Sum rules of spin resolved structure factor \texorpdfstring{$\bm{S^{\alpha,\beta}(q)}$}{Sab(q)} for the Halperin\texorpdfstring{$-(m,m,n)$}{-(m,m,n)} state}
\label{ssec: sum_rules_spin_resolved_Sq}
Using \cref{eq:structure-factor1}, the $q{\to}0$ expansion of $S^{\alpha,\beta}(q)$ is
\begin{align}
    S^{\alpha,\beta}(q)
    &= s^{\alpha,\beta}_0
    + s^{\alpha,\beta}_2 q^2
    + \mathcal{O}(q^4),
    \label{eq:sabexpansion} \\
    s^{\alpha,\beta}_0
    &= \delta_{\alpha,\beta}\frac{\rho_{0,\alpha}}{\rho_0}
    +\frac{\rho_{0,\alpha}\rho_{0,\beta}}{\rho_0}
    \mathcal I^{\alpha,\beta}_0,\\
    s^{\alpha,\beta}_2
    &= -\frac{\rho_{0,\alpha}\rho_{0,\beta}}{\rho_0}
    \frac{\mathcal I^{\alpha,\beta}_2}{4}.
\end{align}
For the Halperin-$(m,m,n)$ state, with $\rho_{0,\up}{=}\rho_{0,\dn}{=}\rho_0/2$, the coefficients
$s^{\alpha,\beta}_0$ and $s^{\alpha,\beta}_2$ were obtained from the OZ analysis of the corresponding two-component Coulomb plasma~\cite{Forrester84, Gail08} as
\begin{subequations}
\label{eq:sab_coeffs}
\begin{align}
s^{\alpha,\beta}_0 &= 0,
\qquad \forall\,\alpha,\beta,\\
s^{\alpha,\beta}_2
&=
\frac{1}{4(m-n)}
\begin{pmatrix}
m & -n\\
-n & m
\end{pmatrix}_{\alpha,\beta}.
\label{eq:sab_2}
\end{align}
\end{subequations}

The condition $s^{\alpha,\beta}_0{=}0$ follows from charge neutrality of the
plasma and is equivalent to the asymptotic behavior $g^{\alpha,\beta}(r){\rightarrow}1$ as $r{\rightarrow}\infty$. The coefficient $s^{\alpha,\beta}_2$ is determined by the long-range Coulomb
interaction of the two-component plasma. In \cref{sub:field-theory}, we present an alternate derivation of \cref{eq:sab_coeffs} using the $K$-matrix Chern-Simons-Maxwell effective field theory.

Unlike the total $S(q)$, exact sum rules for the higher-order coefficients $s_n^{\alpha,\beta}$ ($n\ge4$) are not known. One approach is to diagonalize the two-component plasma~\cite{Gail08} into charge and spin sectors with effective
densities
\begin{equation}
\rho_{0,c}=\rho_{0,\up}+\rho_{0,\dn},
\qquad
\rho_{0,s}=\rho_{0,\up}-\rho_{0,\dn}.
\end{equation}
However, for the spin-singlet Halperin-$(m,m,m{-}1)$ states, $\rho_{0,\up}{=}\rho_{0,\dn}$, so that $\rho_{0,s}{=}0$. Thus, the spin sector corresponds to a plasma of vanishing density, for which the standard one-component plasma sum rules cannot be applied. Consequently, this diagonalization does not provide a route to determine the coefficients $s_n^{\alpha,\beta}$ ($n\ge4$), and to the best of our knowledge, no exact sum rules for these higher-order coefficients are currently available.

\section{Long-wavelength structure factor from the \texorpdfstring{$K$}{K}-matrix Maxwell-Chern-Simons theory}
\label{sub:field-theory}
In this section, we derive the long-wavelength expansion of the $S^{\alpha,\beta}(q)$ within the $K$-matrix Chern-Simons theory~\cite{Wen1990, Wen92a, Wen92b, Wen95}. Throughout this section, we set $\hbar=1$, and use $\delta_{ij}$ for the spatial metric, and employ the Einstein summation convention, where repeated indices are summed over. We will keep our derivations general for $\mathcal{N}$ flavors, with the flavor indices $\mathsf I,\mathsf J= 1,\ldots,\mathcal {N}$ labeling the different components, and then specialize to the $\mathcal{N}=2$ case of our interest. The topological Chern--Simons theory introduced in~\cite{Wen1990, Wen92a, Wen92c, Wen95}, however, is insufficient for describing dynamical properties such as the dynamical and static structure factors. Since the pure Chern--Simons theory contains no propagating bulk degrees of freedom, it must be supplemented with gauge-invariant Maxwell terms to describe the structure factor. The resulting effective theory is the Maxwell--Chern--Simons theory~\cite{Dunne}. The $K$-matrix, $K_{\mathsf I\mathsf J}$ is real symmetric, the effective permittivity matrix $(\varepsilon_e)_{\mathsf I\mathsf J}$ is real symmetric positive definite and and effective permeability $(\mu^{-1})_{\mathsf I\mathsf J}$ is real symmetric positive semidefinite. Unlike $K$, which depends on the universality class of the underlying topological phase, $(\varepsilon_e)$ and $(\mu^{-1})$ are non-universal, depending on the microscopic details. The filling factor of the FQH state is given by $\nu = \mathbf t^T K^{-1} \mathbf t$, where $\mathbf t$ is the charge vector with real elements; correspondingly, the density is $\rho_0 $~\cite{Wen92a, Wen92b, Wen95}.

The low-energy dynamics of the FQH state are described by the $K$-matrix Maxwell-Chern-Simons (KMCS) theory  of the emergent gauge fields $a_\mu^{\mathsf I}$,  and are described by the Lagrangian density
\begin{align}
\mathcal{L}
=
\frac{K_{\mathsf I\mathsf J}}{4\pi}\epsilon^{\mu\nu\rho}a^{\mathsf I}_\mu\partial_\nu a^{\mathsf J}_\rho
+\frac{(\varepsilon_e)_{\mathsf I\mathsf J}}{2}e_i^{\mathsf I}e_i^{\mathsf J}
-\frac{(\mu^{-1})_{\mathsf I\mathsf J}}{2}b^{\mathsf I}b^{\mathsf J},
\label{eq:action}
\end{align}
where $e_i^{\mathsf I}=\partial_0a_i^{\mathsf I}-\partial_i a_0^{\mathsf I}$ and $b^{\mathsf I}=\epsilon_{ij}\partial_i a_j^{\mathsf I}$ are the emergent electric and magnetic fields, respectively. The particle density is related to the emergent magnetic field through $\rho_{\mathsf I}=\mathbf{t}_{\mathsf I} b^{\mathsf I}/(2\pi)$~\cite{Wen92a}.

\subsection{Density response function}
To evaluate the density-density correlation function, we work in the Coulomb gauge, $q_i a_i^{\mathsf I}{=}0$. In two spatial dimensions, the gauge field has a single transverse component, $a_i^{\mathsf I}{=}\hat t_i a_T^{\mathsf I}$ with $\hat t_i{=}\epsilon_{ij}\hat q_j$, giving $b^{\mathsf I}{=}{-}\iota q\,a_T^{\mathsf I}$ and $e_i^{\mathsf I}e_i^{\mathsf J}{=}\omega^2a_T^{\mathsf I}a_T^{\mathsf J}+q^2a_0^{\mathsf I}a_0^{\mathsf J}$. In this gauge, the spatial Chern-Simons term vanishes identically, while the temporal part reduces to $(K_{\mathsf I\mathsf J}/2\pi)a_0^{\mathsf I}b^{\mathsf J}=-(\iota q/2\pi)K_{\mathsf I\mathsf J}a_0^{\mathsf I}a_T^{\mathsf J}$.

Introducing the field doublet notation $\Phi_{\mathsf I}{=}(a_0^{\mathsf I},a_T^{\mathsf I})$, the quadratic action can be written as 
\begin{equation}
  S=\frac12\int\frac{\dd\omega\,\dd^2 \bm q}{(2\pi)^3}\Phi^\dagger\mathcal K\Phi,  
\end{equation}
with kernel is a $2\mathcal{N}\times 2 \mathcal{N}$ matrix
\begin{equation}
\mathcal K(\omega,q)=
\begin{pmatrix}
\varepsilon_e q^2 &
-\dfrac{\iota q}{2\pi}K\\[2mm]
\dfrac{\iota q}{2\pi}K &
\varepsilon_e\omega^2-\mu^{-1}q^2
\end{pmatrix}.
\label{eq:kernel}
\end{equation}

As the action is quadratic, the transverse gauge-field propagator can be computed and is (see Appendix~\ref{app:gauge-propagator})
\begin{align}
\langle a_T^{\mathsf I}a_T^{\mathsf J}\rangle(\omega,q)
=
\iota 
\left[
\varepsilon_e\omega^2
-\mu^{-1}q^2
-\frac{1}{(2\pi)^2}
K\varepsilon_e^{-1}K
\right]^{-1}_{\mathsf I\mathsf J}.
\label{eq:aTaT}
\end{align}
Equation~\eqref{eq:aTaT} shows that the Chern-Simons coupling contributes to the transverse propagator only through the induced term $K\varepsilon_e^{-1}K$. In the absence of the Maxwell term, the transverse propagator vanishes identically, demonstrating that the pure Chern--Simons theory contains no propagating gauge-field degrees of freedom and describes only topological properties.

\subsection{Collective magnetoplasmon modes}
The poles of the transverse gauge-field propagator determine the dispersion of the collective magnetoplasmon modes,
\begin{align}
    \omega_i^2(q)
    =
    \mathcal E_i\!\left[\mathcal W(q)\right],
    \label{eq:magnetoplasmon_energy}
\end{align}
where $\mathcal E_i[\cdot]$ denotes the $i^{\rm th}$ eigenvalue of the enclosed matrix. Here,
\begin{align}
    C(q)
    &=
    \mu^{-1}q^2
    +
    \frac{1}{(2\pi)^2}
    K\varepsilon_e^{-1}K,
    \label{eq:Cq}
    \\
    \mathcal W(q)
    &=
    \varepsilon_e^{-1}C(q)
    =
    \varepsilon_e^{-1}\mu^{-1}q^2
    +
    \left(
        \frac{\varepsilon_e^{-1}K}{2\pi}
    \right)^2.
\end{align}
The corresponding eigenvectors $\{\bm v^{(i)}\}_{i=1}^{\mathcal N}$ determine the linear combination of density fluctuations associated with the $i^{\rm th}$ collective mode, \(\rho^{(i)}{=}v^{(i)}_{\mathsf I} \rho_{\mathsf I}\). Thus, an $\mathcal N$-component system supports $\mathcal N$ branches of gapped magnetoplasmon excitations. In the long-wavelength limit, the gap of the $i^{\rm th}$ magnetoplasmon branch is
\begin{align}
    \Delta_i^{\mathrm{KMCS}} = \frac{1}{2\pi} \mathcal E_i \!\left[ \varepsilon_e^{-1}K \right].
\end{align}
The long-wavelength gaps may be obtained from microscopic calculations, such as the SMA, or ascertained from inelastic light-scattering experiments. Since the $K$ matrix is fully fixed by the topological order of the FQH phase, matching the long-wavelength gaps of the magnetoplasmon branches determines the matrix $\varepsilon_e^{-1}$. Once $\varepsilon_e^{-1}$ is known, the leading small-$q$ dispersion of the magnetoplasmons provides an estimate of $\mu^{-1}$.

Using $\rho_{\mathsf I}=\mathbf{t}_{\mathsf I} b^{\mathsf I}/(2\pi)$ and $\langle b^{\mathsf I}b^{\mathsf J}\rangle=q^2\langle a_T^{\mathsf I}a_T^{\mathsf J}\rangle$, the dynamical density-density correlator is
\begin{align}
\langle\rho_{\mathsf I}\rho_{\mathsf J}\rangle(\omega,q)
= \iota\frac{q^2 \mathbf{t}_{\mathsf I} \mathbf{t}_{\mathsf J}}{(2\pi)^2}
\left[
\varepsilon_e\omega^2-C(q)
\right]^{-1}_{\mathsf I\mathsf J}.
\label{eq:rhorho}
\end{align}

\subsection{Static structure factor}
The $S(q)$ is obtained by integrating the dynamical structure factor over frequency (see Appendix~\ref{app:frequency-integral}),
\begin{align}
S^{\mathsf I,\mathsf J}(q)
=
\frac{1}{\rho_0}\int\frac{\dd\omega}{2\pi}
\langle\rho_{\mathsf I}\rho_{\mathsf J}\rangle(\omega,q),
\label{eq:Sdef}
\end{align}
\begin{align}
S^{\mathsf I,\mathsf J}(q)
=
\frac{q^2 \mathbf{t}_{\mathsf I} \mathbf{t}_{\mathsf J}}{2\rho_0(2\pi)^2}
\left[
\varepsilon_e^{-1/2}
\left(
\varepsilon_e^{-1/2}C(q)\varepsilon_e^{-1/2}
\right)^{-1/2}
\varepsilon_e^{-1/2}
\right]_{\mathsf I\mathsf J}.
\label{eq:Sexact}
\end{align}
Expanding \cref{eq:Sexact} for small $q$ gives the universal leading-order result (see Appendix~\ref{app:frequency-integral})
\begin{equation}
S^{\mathsf I,\mathsf J}(q)
=
\frac{(q\ell)^2 \mathbf{t}_{\mathsf I} \mathbf{t}_{\mathsf J}}{2\nu}
(K^{-1})_{\mathsf I\mathsf J}
+\mathcal O(q^4).
\label{eq:Sresult}
\end{equation}

For Laughlin and Halperin states, Eq~\eqref{eq:Sresult} agrees with the result obtained previously from the plasma analogy \cref{eq:sab_coeffs}~\cite{Forrester84, Gail08}. Summing over all components, we recover the universal long-wavelength coefficient of $S(q)$, i.e., 
$S(q)=\sum_{\mathsf I,\mathsf J}S^{\mathsf I,\mathsf J}(q)=(q\ell)^2/2+\mathcal{O}(q^4)$,
in agreement with the well-known universal result that $s_2=1/2$.

A few remarks are in order. Although the Maxwell coefficients $\varepsilon_e$, $\mu^{-1}$are necessary to obtain a nonzero equal-time density correlator, they cancel in the leading-order static structure factor. Consequently, the leading $\mathcal{O}(q^2)$ coefficient is universal: it is determined entirely by the topological $K$ matrix through $K^{-1}$ and is independent of the non-universal Maxwell parameters.

Finally, we note that $S^{\mathsf I,\mathsf J}(q)$ must be positive semidefinite. Indeed, for an arbitrary real vector $v_{\mathsf I}$,
\begin{equation}
v_{\mathsf I} S^{\mathsf I,\mathsf J}(q)v_{\mathsf J}
\sim
\left\langle
\left(v_{\mathsf I}\rho_{\mathsf I}(\mathbf q)\right)^\dagger
\left(v_{\mathsf J}\rho_{\mathsf J}(\mathbf q)\right)
\right\rangle
\ge0.
\end{equation}
Since $S^{\mathsf I,\mathsf J}(q)=(q\ell)^2\mathbf{t}_{\mathsf I}\mathbf{t}_{\mathsf J}(K^{-1})_{\mathsf I\mathsf J}/(2\nu)+\mathcal O(q^4)$, the leading long-wavelength $S^{\mathsf I,\mathsf J}(q)$ is positive semidefinite only if the $K$ matrix is itself positive definite, i.e., for chiral states.


\subsection{Application to Halperin\texorpdfstring{$\bm{-(m,m,n)}$}{-(m,m,n)} states}
\label{sub:application-to-Halperin}
For the Halperin $(m,m,n)$ state, the $K$ matrix is given by Eq.~\eqref{eq:Halperin-Kmatrix}, and the charge vector is given by $\mathbf{t}=(1,1)$ ~\cite{Wen92c}. Assuming layer-exchange symmetry and $m>n$, $\varepsilon_1 > \varepsilon_2$ and $\mu_1^{-1} >\mu_2^{-1}$, we write the following matrices
\begin{align}
    \varepsilon_e
    &=
    \begin{pmatrix}
        \varepsilon_1 & \varepsilon_2\\
        \varepsilon_2 & \varepsilon_1
    \end{pmatrix},
&
    \mu^{-1}
    &=
    \begin{pmatrix}
        \mu_1^{-1} & \mu_2^{-1}\\
        \mu_2^{-1} & \mu_1^{-1}
    \end{pmatrix}.
\end{align}
The theory supports two collective magnetoplasmon branches corresponding to the symmetric (charge) and antisymmetric (spin) density fluctuations,
\begin{align}
    \rho_c &= \rho_1+\rho_2,
&
    \rho_s &= \rho_1-\rho_2.
\end{align}

The charge mode has dispersion
\begin{align}
    \omega_c^2(q)
    =
    \frac{q^2}{\varepsilon_1+\varepsilon_2}
    \left(
        \frac1{\mu_1}
        +
        \frac1{\mu_2}
    \right)
    +
    \left[
        \frac{m+n}
        {2\pi(\varepsilon_1+\varepsilon_2)}
    \right]^2,
\end{align}
with gap
\begin{align}
    \Delta_c^{\mathrm{KMCS}}
    =
    \frac{m+n}
    {2\pi(\varepsilon_1+\varepsilon_2)}.
    \label{eq:eft_gap_c}
\end{align}

Similarly, the spin mode's dispersion is
\begin{align}
    \omega_s^2(q)
    =
    \frac{q^2}{\varepsilon_1-\varepsilon_2}
    \left(
        \frac1{\mu_1}
        -
        \frac1{\mu_2}
    \right)
    +
    \left[
        \frac{m-n}
        {2\pi(\varepsilon_1-\varepsilon_2)}
    \right]^2,
\end{align}
with gap
\begin{align}
    \Delta_s^{\mathrm{KMCS}}
    =
    \frac{m-n}
    {2\pi(\varepsilon_1-\varepsilon_2)}.
      \label{eq:eft_gap_s}
\end{align}
The relative ordering of the charge and spin gaps is controlled by the $\varepsilon_e$ matrix: $\Delta_c^{\mathrm{KMCS}} {>} \Delta_s^{\mathrm{KMCS}}$ for $\varepsilon_1 {>} (m/n)\varepsilon_2$, while $\Delta_c^{\mathrm{KMCS}} {<} \Delta_s^{\mathrm{KMCS}}$ for $\varepsilon_1 {<} (m/n)\varepsilon_2$. Thus, for the spin-singlet Halperin \((m,m,m-1)\) states, the spin mode is the lower-energy neutral excitation.

The inverse $K$ matrix is
\begin{align}
    K^{-1}
    =
    \frac{1}{m^2-n^2}
    \begin{pmatrix}
        m & -n\\
        -n & m
    \end{pmatrix},
\end{align}
from which the long-wavelength flavor-resolved static structure factor follows,
\begin{align}
        S^{\mathsf I,\mathsf J}(q)
    =
    \frac{(q\ell)^2}{4(m-n)}
    \begin{pmatrix}
        m & -n\\
        -n & m
    \end{pmatrix}_{\mathsf I\mathsf J}.
\end{align}
This agrees with Eq.~\eqref{eq:sab_2}, obtained independently from the OZ equation.

\section{Density-wave modes}
\label{sec:collective}
In this section, we present the SMA gaps for the symmetric density-wave (SDW), antisymmetric density-wave (ASDW) modes of spin-singlet FQH states and spin-flip density-wave (SFDW) mode for spin-polarized FQH states.

\subsection{Symmetric density wave}
GMP~\cite{Girvin85, Girvin86} proposed a variational ansatz for the neutral excitation that is obtained by acting the LLL-projected density operator on the ground state $\ket{\Psi_0}$:
\begin{align}
    \ket{\Psi_q^{\mathrm{SDW}}} = \bar\rho_q^{I} \ket{\Psi_0}. \label{eq:psi_SDW}
\end{align}
The excitation energy associated with this SDW mode is
\begin{align}
    \Delta^{\mathrm{SDW}}(q) = \frac{ \expval{ \bar\rho_{-q}^{I}\, \bar H\, \bar\rho_q^{I} }{\Psi_0} }{ \expval{ \bar\rho_{-q}^{I} \bar\rho_q^{I} }{\Psi_0} } - \expval{\bar H}{\Psi_0},
    \label{eq:SDW-GAP0}
\end{align}
where $\ket{\Psi_0}$ is assumed to be a normalized eigenstate of $\bar H$, i.e., $\braket{\Psi_0}=1$ and $\bar{H}\ket{\Psi_0}=E_{0}\ket{\Psi_0}$, with $E_0$ being the ground state energy, and the projected density operator satisfies $(\bar\rho_q^{I})^\dagger=\bar\rho_{-q}^{I}$.

Using these properties, the excitation gap can be rewritten in terms of a single commutator,
\begin{align}
    \Delta^{\mathrm{SDW}}(q) = \frac{ \expval{ \bar\rho_{-q}^{I} \left[ \bar H, \bar\rho_q^{I} \right] }{\Psi_0} }{ \expval{ \bar\rho_{-q}^{I} \bar\rho_q^{I} }{\Psi_0} }.
\end{align}
The denominator is the projected static structure factor,
\begin{align}
    \bar S(q) = \frac{1}{N} \expval{ \bar\rho_{-q}^{I} \bar\rho_q^{I} }{\Psi_0},
\end{align}
which is related to $S(q)$ through
$
    \bar S(q) = S(q) - \left(1-e^{-q^2/2} \right)
$
(see \cref{app:Relation between the projected and unprojected structure factors})~\cite{Girvin86}.

The numerator in \cref{eq:SDW-GAP0} is the projected oscillator strength,
\begin{align}
    \bar F(q) = \frac{1}{N} \expval{ \bar\rho_{-q}^{I} \left[ \bar H, \bar\rho_q^{I} \right] }{\Psi_0}.
\end{align}
For a translationally and rotationally invariant ground state, $\bar F(q)$ is an even function of $q$,
$
    \bar F(q) = \bar F(-q).
$
This allows $\bar F(q)$ to be expressed in the more convenient double-commutator form~\cite{Girvin86},
\begin{align}
    \bar F(q) = \frac{1}{2N} \expval{ \left[ \bar\rho_{-q}^{I}, \left[ \bar H, \bar\rho_q^{I} \right] \right] }{\Psi_0}.
\end{align}
Thus, the SDW gap is
\begin{align}
    \Delta^{\mathrm{SDW}}(q) = \frac{\bar F(q)} {\bar S(q)} =
    \frac12\frac{ \expval{ \left[ \bar\rho_{-q}^{I}, \left[ \bar H, \bar\rho_q^{I} \right] \right] }{\Psi_0} }{ \expval{ \bar\rho_{-q}^{I} \bar\rho_q^{I} }{\Psi_0} },
    \label{eq:SDW-gap}
\end{align}
which is referred to as the GMP gap equation~\cite{Girvin86}. Using the commutation relations in \cref{eq:spinful-commutator-algebra} together with $\bar{H}$ of \cref{eq:projected-Hamiltonian}, $\bar F(q)$ can be evaluated explicitly, yielding
\begin{align}
    \bar F(q)
    =&\, 2 \int \frac{\dd^2 \bm k}{(2\pi)^2} \, v(k)\, \swedgesq{k}{q} \nonumber\\
    &\times \left[ e^{\bm q\cdot\bm k} \bar S(\abs{\bm k+\bm q}) - e^{-q^2/2} \bar S(k)
    \right].
    \label{eq:oscillator-strength}
\end{align}
Expanding $\bar F(k)$ and $\bar S(k)$ up to $\mathcal{O}(k^4)$~\cite{Giuliani08}, results in the long-wavelength SDW gap
\begin{align}
 \Delta^{\mathrm{SDW}}(q\to0)
&=
\int_0^\infty \dd k\, \frac{k^2\bar S(k)}{64\pi (s_4 + 1/8)} \times \nonumber \\
& \left[ k(k^2-4)v(k) +\left(3-2k^2\right)v'(k) + k v''(k) \right].
\label{eq:k1_SDW_gap}
\end{align}
For Coulomb interaction $v(q) = 2\pi/q$, resulting in
\begin{align} 
      \Delta^{\mathrm{SDW,\, C}}(q\to0) = 
    \frac{1}{32}
    \int_0^{\infty} \dd k \frac{\bar S(k)(k^4 - 2k^2 -1)}{s_4 + 1/8}.
    \label{eq:k0_SDW_gap}
\end{align}

A few comments are in order: the derivation of the above SDW gap assumes that $\ket{\Psi_{0}}$ is an exact eigenstate of $\bar{H}$. Nevertheless, we are going to use the same formalism but employ trial wavefunctions for two reasons: 1) The exact eigenstates for $\bar{H}$ can only be constructed for small systems via exact diagonalization, and 2) the trial wavefunctions we use provide excellent representations of the exact Coulomb eigenstates for small systems~\cite{Wu93, Balram15a, Liu20}, and have the added advantage that they can be evaluated for fairly large systems allowing us to make more reliable estimates for the thermodynamic limits. As the chosen ansatz provides a very accurate representation of the actual eigenstate, the gap obtained from the SDW gap equation using the $\bar{S}(q)$ of the chosen ansatz is expected to provide a good approximation to the SDW gap.

\subsection{Anti-symmetric density wave}

Similar to the GMP construction, an anti-symmetric density-wave (ASDW) excitation for spinful FQH states is obtained by applying the $z$-spin-density operator on the ground state~\cite{Rasolt86, Dora25},
\begin{align}
    \ket{\Psi_q^{\mathrm{ASDW}}}
    =
    \bar\rho_q^{z}\ket{\Psi_0},
    \label{eq:psi_ASDW}
\end{align}
which corresponds to an out-of-phase density fluctuation of the two spin components since $\bar\rho_q^{z}=\bar\rho_q^{\uparrow}-\bar\rho_q^{\downarrow}$. Following the same steps as for the SDW mode, the ASDW gap is
\begin{align}
    \Delta^{\mathrm{ASDW}}(q)
    =
    \frac{\bar F^z(q)}{\bar S^z(q)}
    =
    \frac{1}{2}
    \frac{
    \expval{\left[\bar\rho_{-q}^{z},
    \left[\bar H,\bar\rho_q^{z}\right]\right]}{\Psi_0}}
    {\expval{\bar\rho_{-q}^{z}\bar\rho_q^{z}}{\Psi_0}},
    \label{eq:ASDW-gap}
\end{align}
where
\begin{align}
    \bar S^z(q)
    &=
    \frac{1}{N}
    \expval{\bar\rho_{-q}^{z}\bar\rho_q^{z}}{\Psi_0},
    \\
    \bar F^z(q)
    &=
    \frac{1}{2N}
    \expval{\left[\bar\rho_{-q}^{z},
    \left[\bar H,\bar\rho_q^{z}\right]\right]}{\Psi_0}.
\end{align}
Analogous to $\bar{F}(k)$, its $z$-version is
\begin{align}
    \bar F^z(q)
    =&\,
    2
    \int
    \frac{\dd^2\bm k}{(2\pi)^2}
    \,
    v(k)\,
    \swedgesq{k}{q}
    \nonumber\\
    &\times
    \left[
    e^{\bm q\cdot\bm k}
    \bar S^z(|\bm k+\bm q|)
    -
    e^{-q^2/2}
    \bar S(k)
    \right].
\end{align}
The quantity $\bar{S}^{z}(q)$ is related to $\bar{S}^{\alpha, \beta}(q)$ through
\begin{align}
    \bar S^z(q)
    &=
    \bar S^{\uparrow\uparrow}(q)
    +
    \bar S^{\downarrow\downarrow}(q)
    -
    \bar S^{\uparrow\downarrow}(q)
    -
    \bar S^{\downarrow\uparrow}(q).
\end{align}
For spin-singlet states,
\begin{align}
    \bar S^z(q)
    =
    2\left[
    \bar S^{\uparrow\uparrow}(q)
    -
    \bar S^{\uparrow\downarrow}(q)
    \right].
    \label{eq:szbar}
\end{align}
Finally, $\bar S^z(k)$ and $S^z(k)$ are related by
\begin{align}
    \bar S^z(q)
    =
    S^z(q)
    -
    \left(1-e^{-q^2/2}\right),
\end{align}
as shown in \cref{app:Relation between the projected and unprojected structure factors}.

The $k\to0$ ASDW gap is computed by expanding $\bar F^z(k)$ and $\bar S^z(k)$ upto $\mathcal{O}(k^2)$ and using \cref{eq:szbar} to get
\begin{align} 
    \Delta^{\mathrm{ASDW}}(q\to0)
    = 
    -\int_0^\infty  \dd k \frac{ \,k^3 v(k)\, S^{\up,\dn}(k)}{4\pi (s^{\up,\up}_2 - s^{\up,\dn}_2 - 1/4)} .
    \label{eq:k0_ASDW_gap}
\end{align} 

\subsection{Spin-flip density wave}

Another density-wave excitation is the spin-flip density-wave (SFDW) mode, is one in which a particle is moved to the opposite spin branch. We consider this mode only for fully polarized states since for $SU(2)$ symmetric interactions that we focus on, for singlet states the SFDW mode and ASDW mode have identical dispersions as they belong to different $\mathbb{S}_z$ sectors of the same $\mathbb{S}=1$ multiplet (SFDW has $\mathbb{S}_z=\pm 1$, and ASDW has $\mathbb{S}_z=0$). Within the SMA, the corresponding variational state is constructed by acting with the spin-lowering projected density operator on the polarized ground state,
\begin{align}
    \ket{\Psi_q^{\mathrm{SFDW}}}
    =
    [\bar\rho_q]^{-}\ket{\Psi_0},
\end{align}
where
\begin{align}
    [\bar\rho_q]^{-}
    =
    e^{-q^2/4}
    \sum_{i=1}^{N}
    \sigma_i^{-}\otimes\tau_q(i),
\end{align}
with $\sigma^-=\sigma_x-i\sigma_y$. This operator flips the spin of a particle while imparting it momentum $k$. Its Hermitian conjugate satisfies $([\bar\rho_q]^{-})^\dagger=[\bar\rho_{-q}]^{~+}$.

Proceeding exactly as for the SDW and ASDW modes, the SFDW gap is
\begin{align}
    \Delta^{\mathrm{SFDW}}(q)
    =
    \frac12
    \frac{
    \expval{
    \left[
    [\bar\rho_{-q}]^{+},
    \left[
    \bar H,
    [\bar\rho_q]^{-}
    \right]
    \right]
    }{\Psi_0}
    }
    {
    \expval{
    [\bar\rho_{-q}]^{+}
    [\bar\rho_q]^{-}
    }{\Psi_0}
    }.
\end{align}

Since the ground state is fully spin polarized, $\ket{\Psi_0}=\Psi(\{z_i\})\ket{\uparrow\uparrow\cdots\uparrow}$, the action of the spin-raising operator is trivial, $\bar\rho_q^{+}\ket{\Psi_0}=0$. Furthermore,
\begin{align}
    \left[
    \bar\rho_{-q}^{+},
    [\bar\rho_q]^{-}
    \right]
    =
    2e^{-q^2/2}\mathbb S^z,
\end{align}
where $2\mathbb S^z=\sum_i\sigma_i^z=\bar\rho_0^I$. It follows immediately that
\begin{align}
    \expval{
    \bar\rho_{-q}^{+}
    [\bar\rho_q]^{-}
    }{\Psi_0}
    =
    Ne^{-q^2/2}.
\end{align}

Using the algebra of projected operators, the SFDW oscillator strength can be evaluated explicitly, yielding, up to the Zeeman splitting,
\begin{align}
    \Delta^{\mathrm{SFDW}}(q)
    =
    \frac{2}{(2\pi)^2}
    \int
    \dd^2\bm k\,
    v(k)
    \sin^2\!\left(\frac{k\wedge q}{2}\right)
    \left[1-S(k)\right].
\end{align}

The angular integration can be performed analytically, leading to the well-known expression~\cite{Rasolt86, Nakajima94, Dora25}
\begin{align}
    \Delta^{\mathrm{SFDW}}(q)
    =
    \int
    \frac{\dd k}{2\pi}
    \,k\,
    v(k)
    \left[1-J_0(qk)\right]
    \left[1-S(k)\right],
    \label{eq:SFDW-gap}
\end{align}
where $J_0(x)$ is the Bessel function of the first kind.

For the integer quantum Hall state at $\nu=1$, using the Coulomb interaction together with $S_{\nu=1}(q)=1-e^{-q^2/2}$, one recovers the exact result of Kallin and Halperin~\cite{Kallin84},
\begin{align}
\label{eq: SFDW_nu_1}
    \Delta_{\nu=1}^{\mathrm{SFDW}}(q)
    =
    \sqrt{\frac{\pi}{2}}
    \left[
    1-
    e^{-q^2/4}
    I_0\!\left(\frac{q^2}{4}\right)
    \right],
\end{align}
where $I_0(x)$ is the modified Bessel function of the first kind. The same approach also yields the exact SFDW dispersion in higher Landau levels, wherein one replaces the interaction $v(q)$ with the effective interaction $v(q)\rightarrow v(q)[L_n(q^2/2)]^2$, where $L_n(x)$ is the Laguerre polynomial for the $n$th Landau level.

In the $k\to0$ limit, the SFDW mode dispersed quadratically with $k$, i.e., $\Delta^{\mathrm{SFDW}}(q)=D_s q^2$, up to the Zeeman splitting, where the spin stiffness is
\begin{align}
    D_s
    =
    \int
    \frac{\dd k}{8\pi}
    \,k^3
    v(k)
    \left[1-S(k)\right].
    \label{eq:spin-stiffness}
\end{align}
Expanding Eq.~\eqref{eq: SFDW_nu_1} at small-$q$ gives $D^{\nu=1}_s=\sqrt{2\pi}/8$.

\section{Fitting pair correlation}
\label{sec:fitting}
In this section, we describe the parameterization of the $g^{\alpha, \beta}(r)$ that enforces the exact sum-rule constraints that we derived in~\cref{ssec: sum_rules_Sq} and~\cref{ssec: sum_rules_spin_resolved_Sq}. We first review Girvin's nonorthogonal basis for expanding $g(r)$~\cite{Girvin84a} and then discuss how it can be improved upon by using the orthogonal basis introduced by Fulsebakke \textit{et al}.~\cite{Fulsebakke23}.

\subsection{Girvin's non-orthogonal basis} 
Girvin~\cite{Girvin84a} proposed the following convenient parametrization of the $g(r)$ of a FQH state
\begin{align}
g(r) &= 1 - e^{-r^2/2}
+ \sum_{j=1}^{\infty} c_j \mathfrak h_j(r), \label{eq:gcorr}\\
g^{\alpha,\beta}(r) &= 1 - e^{-r^2/2}
+ \sum_{j=1}^{\infty} c^{\alpha,\beta}_j \mathfrak h_j(r), \label{eq:pair-corr-2}\\
\mathfrak h_j(r) &= \frac{2}{j!} \left(\frac{r^2}{4}\right)^j e^{-r^2/4}. \label{eq:pair-corr-3}
\end{align}
For fermionic states, the expansion of $g(r)$ contains only odd values of $j$, reflecting the antisymmetry of the wavefunction. In contrast, $g^{\alpha,\beta}(r)$ generally contains both even and odd values of $j$. Moreover, if $g^{\alpha,\beta}(r)$ is finite at $r=0$, the expansion must also include the $j=0$ term. The short-distance behavior of the $g(r)$ further constrains the expansion coefficients. For example, for the $1/m$ Laughlin state, the $g(r)$ satisfies $g(r)\sim r^{2m}$ as $r\to0$, implying $c_j=-1$ for all $j<m$. 

Substituting \cref{eq:gcorr,eq:pair-corr-2} into \cref{eq:SF_from_pair,eq:structure-factor} yields,
\begin{align}
S(q) &= 1 - \nu e^{-q^2/2}
+ 4\nu \sum_{j=1}^{\infty}
c_j e^{-q^2} L_j(q^2), \label{eq:SF-total}\\
S^{\alpha,\beta}(q)
&= \frac{\delta_{\alpha,\beta}}{2}
- \frac{\nu}{4} e^{-q^2/2}
+ \nu \sum_{j=1}^{\infty}
c^{\alpha,\beta}_j e^{-q^2} L_j(q^2).
\label{eq:SF-component}
\end{align}

The fitting coefficients $\{c_j\}$ are determined by requiring that $S(q)$ reproduces the exact long-wavelength expansion discussed in~\cref{ssec: sum_rules_Sq}. This is achieved by matching the small-$q$ expansion of \cref{eq:SF-total} with \cref{eq:sexpansion} order by order in $q^2$, which mandates that the fitting coefficients satisfy the following rules:
\begin{subequations}
\label{eq:g-constraints-girvin}
\begin{align}
&\sum_{j=1}^{\infty} c_j = \frac{s_0-1}{4\nu}+\frac14, \label{eq:cons1}\\
&\sum_{j=1}^{\infty} (j+1)c_j = \frac18-\frac{s_2}{4\nu}, \label{eq:cons2}\\
&\sum_{j=1}^{\infty} (j+1)(j+2)c_j = \frac{s_4}{\nu}+\frac18, \label{eq:cons3}\\
&\sum_{j=1}^{\infty} (j+1)(j+2)(j+3)c_j = \frac{3}{16}-\frac{9s_6}{\nu}. \label{eq:cons4}
\end{align}
\end{subequations}

Similarly, for $S^{\alpha, \beta}$, comparing \cref{eq:SF-component} with the $q{\to}0$ expansion in \cref{eq:sabexpansion} order by order in $q^2$ yields
\begin{subequations}
\label{eq:constraint-gab}
\begin{align}
&\sum_{j=1}^{\infty} c_j^{\alpha,\beta}
=
\frac{s_0^{\alpha,\beta}}{\nu}
-\frac{\delta_{\alpha,\beta}}{2\nu}
+\frac12,\\
&\sum_{j=1}^{\infty} (j+1)c_j^{\alpha,\beta}
=
\frac18-\frac{s_2^{\alpha,\beta}}{\nu}.
\end{align}
\end{subequations}

The coefficients $\{c_j\}$ and $\{c_j^{\alpha,\beta}\}$ are obtained by fitting the numerical data for $g(r)$ and $g^{\alpha,\beta}(r)$ while simultaneously enforcing the constraints in \cref{eq:g-constraints-girvin} and \cref{eq:constraint-gab}, respectively. In practice, the expansion is truncated at a finite order $n_{\rm max}$, and the coefficients with $j\leq n_{\rm max}$ are treated as fitting parameters.

Once the fitting coefficients have been determined, the per-particle Coulomb energy, the spin stiffness, and the long wavelength gaps can be evaluated using the following formulas
\begin{subequations}
\label{eq:girvin-fitting-observables}
\begin{align}
E^{\mathrm{C}} &= -\nu\sqrt{\frac{\pi}{8}}
+ \nu \sum_{j} c_j \frac{\Gamma(j+1/2)}{\Gamma(j+1)}, \\
D_s^{\mathrm{C}} &=
\frac{\nu}{4} \sqrt{\frac{\pi}{2}}
-
\frac{\nu}{4} \sum_{j=1}^{\infty} c_j\,
\frac{\Gamma\!\left(j+\frac{1}{2}\right)}
{\Gamma(j+1)(1-2j)}, 
\end{align}
\begin{align}
\Delta^{\mathrm{SDW},\,\mathrm{C}}(q\to0)
&=
\frac{\nu}{16(s_4+1/8)}
\sum_{j=1}^{\infty}
\frac{c_j}{\Gamma(j+1)} \times \nonumber \\
&\quad
\frac{1}{16} (8j (5-2 j)-15)  \Gamma \left(j-\frac{3}{2}\right),\\
    \Delta^{\mathrm{ASDW,\, C}}(q\to0)
    &= 
    \frac{\nu}{8 (s^{\up,\up}_2 - s^{\up,\dn}_2 - 1/4)} \times \nonumber \\
 &\quad\quad
 \Bigg[
  \sqrt{\frac{\pi}{2}} +
 \sum_{j=1}^{\infty}
c^{\up,\dn}_j \frac{\Gamma(j - 1/2)}{2\Gamma(j+1)}
    \Bigg].
\end{align}
\end{subequations}

Care must be exercised, however, when using these expressions since the $\mathfrak{h}_n(r)$ defined in Eq.~\eqref{eq:pair-corr-3} form a set of non-orthogonal basis functions satisfying
\begin{align}
\label{eq: define_M_mn}
\braket{\mathfrak{h}_n}{\mathfrak{h}_m}
&=2\pi\int_0^\infty \dd r\,r\,\mathfrak{h}_n(r)\mathfrak{h}_m(r)
=M_{m,n}\nonumber\\
&=\frac{1}{n!\,m!}\left[\left(\frac{m+n}{2}\right)!\right]^2.
\end{align}
Since the overlap matrix $M_{m,n}$ is non-diagonal and typically ill-conditioned, the fitting coefficients become numerically unstable, leading to a loss of accuracy during the fitting process.

\subsection{Fulsebakke~\emph{et al.}'s orthogonal basis}
To overcome this issue, Fulsebakke~\emph{et al.}~\cite{Fulsebakke23} constructed an orthogonal version of Girvin's basis using the Gram-Schmidt procedure in spherical geometry~\cite{Haldane83} and subsequently took its planar limit. The resulting orthogonal basis for the $g(r)$ is
\begin{align}
g(r)
&=
1-e^{-r^2/2}
+d_0e^{-r^2/2}
+\sum_{n=1}^{\infty}d_nG_n(r),
\label{eq:g-OE}
\\
g^{\alpha,\beta}(r)
&=
1-e^{-r^2/2}
+d_0^{\alpha,\beta}e^{-r^2/2}
+\sum_{n=1}^{\infty}d_n^{\alpha,\beta}G_n(r),
\label{eq:gab-OE}
\\
G_n(r)
&=
\frac{(-1)^nr^2e^{-r^2/2}}{\sqrt{\pi n(n+1)}}
L_{n-1}^{(2)}(r^2),
\label{eq:basisG}
\end{align}
where $L_n^{(s)}(x)$ is the associated Laguerre polynomial. The basis functions $\{G_n(r)\}$ are orthonormal, i.e.,
\[
\braket{G_n}{G_m}
=
2\pi\int_0^\infty \dd r\,r\,G_n(r)G_m(r)
=
\delta_{nm}.
\]
Moreover, these basis states are related to Girvin's basis states through $G_m(r)=\sum_nM_{m,n}\mathfrak{h}_n(r)$, where $M_{m,n}$ is given in \cref{eq: define_M_mn}. The coefficients $\{d_n,d_n^{\alpha,\beta}\}$ vanish as $n\to\infty$, allowing the expansion to be reliably truncated at a sufficiently large value of $n=n_{\rm max}$. The term $d_0e^{-r^2/2}$ is included to describe states with $g(0)\neq0$, for which $d_0=g(0)$. We note that the function $e^{-r^2/2}$ is not orthogonal to the basis functions $\{G_n(r)\}$.

Substituting \cref{eq:g-OE,eq:gab-OE} into \cref{eq:SF_from_pair,eq:structure-factor} results in
\begin{subequations}
\label{eq:s-OE}
\begin{align}
S(q)
&=
1-\nu e^{-q^2/2}
\left(
1-d_0+\sum_{n=1}^{\infty}d_n\mathfrak{I}_n(q)
\right),
\label{eq:S-OE}
\\
S^{\alpha,\beta}(q)
&=
\frac{\delta_{\alpha,\beta}}{2}
-\frac{\nu}{4}e^{-q^2/2}
\left(
1-d_0^{\alpha,\beta}
+\sum_{n=1}^{\infty}
d_n^{\alpha,\beta}\mathfrak{I}_n(q)
\right),
\label{eq:Sab-OE}
\end{align}
\end{subequations}
where
\begin{align}
\mathfrak{I}_n(q)
&=
\sum_{t=0}^{n}\lambda_t^{(n)}q^{2t},
\label{eq:basisJ}
\\
\lambda_t^{(n)}
&=
\frac{(-1)^n}{t!\sqrt{\pi n(n+1)}}
\sum_{s=t}^{n}
(-2)^{s-t}
\binom{n+1}{n-s}
\binom{s}{t}s.
\end{align}

Matching the small-$q$ expansion of \cref{eq:S-OE} with \cref{eq:sexpansion} order by order in powers of $q$ yields
\begin{subequations}
\label{eq:s_coeffs_constraints}
\begin{align}
s_0
&=
1-\nu(\Lambda_0-d_0+1),
\label{eq:s0}
\\
s_2
&=
\frac{\nu}{2}
(\Lambda_0-2\Lambda_1-d_0+1),
\label{eq:s2}
\\
s_4
&=
-\frac{\nu}{8}
(\Lambda_0-4\Lambda_1+8\Lambda_2-d_0+1),
\label{eq:s4}
\\
s_6
&=
\frac{\nu}{48}
(\Lambda_0-6\Lambda_1+24\Lambda_2-48\Lambda_3-d_0+1).
\label{eq:s6}
\end{align}
\end{subequations}
Here $\Lambda_0=\sum_{n=1}^{\infty}d_n\lambda_0^{(n)}$, while for $t>0$,
$\Lambda_t=\sum_{n=t}^{\infty}d_n\lambda_t^{(n)}$. Additionally, the short-distance behavior $g(r)\sim r^{2m}$ as $r\to0$ requires all terms up to $\mathcal{O}(r^{2m-2})$ in \cref{eq:g-OE} to vanish, providing an independent set of linear constraints on the coefficients $\{d_n\}$. Ref.~\cite{Fulsebakke23} implemented only the constraint corresponding to $s_0$ [Eq.~\eqref{eq:s0}], while in this work, we also impose the additional constraints on $s_2$, $s_4$, and $s_6$ [Eqs.~\eqref{eq:s2}-\eqref{eq:s6}].

Similarly, matching the small-$q$ expansion of \cref{eq:Sab-OE} with \cref{eq:sabexpansion} gives
\begin{subequations}
\label{eq:sab_coeffs_constraints}
\begin{align}
s_0^{\alpha,\beta}
&=
\frac{\delta_{\alpha,\beta}}{2}
-\frac{\nu}{4}
(\Lambda_0^{\alpha,\beta}-d_0^{\alpha,\beta}+1),
\label{eq:s0ab}
\\
s_2^{\alpha,\beta}
&=
\frac{\nu}{8}
(\Lambda_0^{\alpha,\beta}
-2\Lambda_1^{\alpha,\beta}
-d_0^{\alpha,\beta}
+1).
\label{eq:s2ab}
\end{align}
\end{subequations}
Here $\Lambda_0^{\alpha,\beta}=\sum_{n=1}^{\infty}d_n^{\alpha,\beta}\lambda_0^{(n)}$, while for $t>0$,
$\Lambda_t^{\alpha,\beta}=\sum_{n=t}^{\infty}d_n^{\alpha,\beta}\lambda_t^{(n)}$.

The coefficients $\{d_n\}$ and $\{d_n^{\alpha,\beta}\}$ are determined by fitting the numerical data for $g(r)$ and $g^{\alpha,\beta}(r)$ while simultaneously enforcing the constraints in \cref{eq:s_coeffs_constraints} and \cref{eq:sab_coeffs_constraints}, respectively. The per-particle Coulomb energy, spin stiffness, and $k{\to}0$ gaps are related to the fitting coefficients $\{d_n\}$ as
\begin{subequations}
\label{eq:orthogonal-fitting-observables}
\begin{align}
E^{\mathrm{C}}
&= 
-\frac{\nu}{2}\left[\sqrt{\frac{\pi}{2}}\left(1-d_0\right) + \sum_{n=1}^{\infty} d_n \sum_{t=0}^{n} 2^{\,t-\frac12} \Gamma\!\left(t+\frac12\right) \lambda_t^{(n)}\right], \\
D_s^C
&=
\frac{\nu}{4} \left[ \sqrt{\frac{\pi}{2}}\left(1-d_0\right) + \sum_{n=1}^{\infty} d_n \sum_{t=0}^{n} 2^{\,t+\frac12} \Gamma\!\left(t+\frac32\right) \lambda_t^{(n)} \right],
\end{align} 
\begin{align} 
&\Delta^{\mathrm{SDW,\,C}}(q\to 0)
=-\frac{ \nu}{16(s_4 + 1/8)} \times \nonumber\\ 
&\quad\quad\sum_{n=1}^{\infty} d_n
\sum_{t=0}^{n}\lambda_t^{(n)} 2^{t + \frac{1}{2}} t (t+1) 
\Gamma \left(t+\frac{1}{2}\right) 
\end{align} 
\begin{align}
&\Delta^{\mathrm{ASDW,\, C}}(q\to 0) = 
\frac{\nu}{8 (s^{\up,\up}_2 - s^{\up,\dn}_2 - 1/4)}\times \\
&\Bigg[ \sqrt{\frac{\pi }{2}}
\left( 1-d_0^{\up,\dn} \right)
+
\sum_{n=1}^{\infty} d_n^{\up,\dn} \sum_{t=0}^{n}
\lambda_t^{(n)} 2^{t+\frac{1}{2}} \Gamma \left(t+\frac{3}{2}\right)\Bigg]. \nonumber
\end{align}
\end{subequations}
Owing to its numerical stability, we prefer to use the orthogonal basis. However, in certain cases, such as $1/5$ Laughlin, we have not been able to obtain a good fit with the orthogonal basis, presumably because of its larger correlation length; therefore, we use the non-orthogonal basis for it.

\begin{figure*}[htpb]
    \centering
    \includegraphics[width=\linewidth]{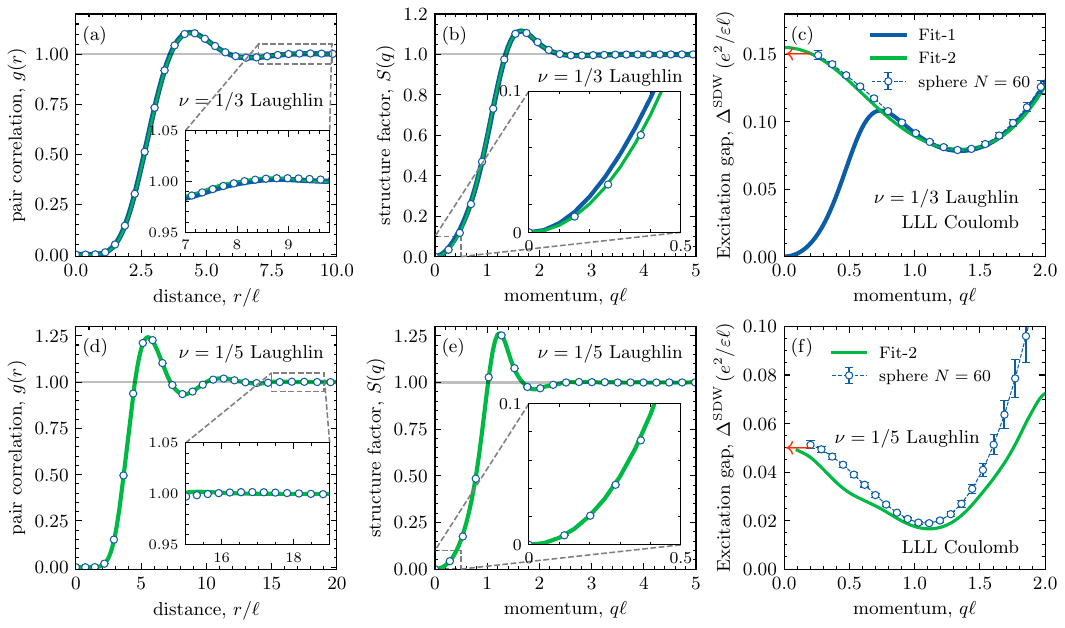}
        \caption{
\textbf{Fits for the density-density correlators of Laughlin states.}
(a) Pair-correlation function $g(r)$ of the $\nu=1/3$ Laughlin state obtained from the orthogonal basis, Eq.~\eqref{eq:g-OE}, compared with Monte Carlo (MC) data on the sphere. Two fitting schemes are considered: {Fit-1}, imposing only the $s_0=0$ constraint, and {Fit-2}, imposing the $s_0$, $s_2$, $s_4$, and $s_6$ constraints, Eq.~\eqref{eq:s_coeffs_constraints}. Inset: long-distance behavior.
(b) Static structure factor $S(q)$ corresponding to the two fits, compared with the MC data. Inset: enlarged view of the long-wavelength regime.
(c) Coulomb GMP gap computed from Eq.~\eqref{eq:SDW-gap}. While Fit-1 fails to reproduce the correct low-$q$ behavior, Fit-2 agrees well with the sphere results in that regime (reproduced from Ref.~\cite{Dora24}). MC error bars in (a) and (b) are smaller than the symbols.
(d)-(f) Same as (a)-(c), but for the $1/5$ Laughlin state, using the Fit-2 scheme with the non-orthogonal basis, Eq.~\eqref{eq:gcorr}. The red arrows in (c) and (f) mark the $q{\to}0$ gap computed using \cref{eq:k0_SDW_gap}.
        }
    \label{fig:Laughlin_fitting}
\end{figure*}

\begin{figure}[htpb]
    \centering
    \includegraphics[width=\linewidth]{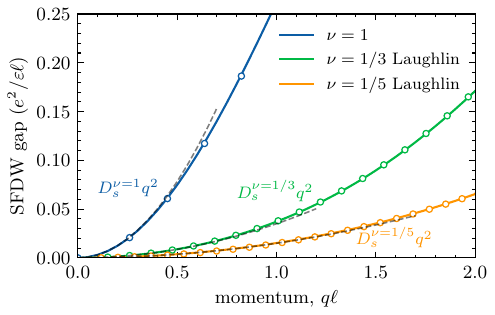}
    \caption{
    \textbf{Spin-flip density-wave (SFDW) gaps.} SFDW gaps for $\nu=1,1/3,1/5$. Solid lines are computed using \cref{eq:SFDW-gap}, with the $\nu=1$ case given by the analytical result in \cref{eq: SFDW_nu_1}. Markers show results obtained on the sphere using the methodology presented in Refs.~\cite{Nakajima94, Dora25} for $N=60$. For $\nu=1/3$ and $1/5$, the statistical error bars from the Monte Carlo simulations are smaller than the markers. Dashed lines show the quadratic fit $D_s q^2$, with $D_s^{\nu=1}=\sqrt{2\pi}/8$, $D_s^{\nu=1/3}=0.035$, and $D_s^{\nu=1/5}=0.015$ all in Coulomb units of $e^2/(\varepsilon\ell)$.
    }
    \label{fig:SFDW-gaps}
\end{figure}

\section{Results: Pair correlations and structure factor}
\label{sec:results-pair-corr-SF}
In this section, we present the $g(r)$ and $S(q)$ obtained using \cref{eq:g-OE,eq:gab-OE} for several FQH states. The data fitted to \cref{eq:g-OE,eq:gab-OE} are generated from Metropolis Monte Carlo (MC) simulations on the sphere~\cite{Haldane83}, with the arc distance on the sphere identified with the planar distance during the fitting procedure. We first benchmark the fitting procedure using the Laughlin state, comparing our results on its $g(r)$, $S(q)$, and GMP dispersion against well-established results on it. We then apply the same approach to obtain new results on spin-singlet FQH states.

\subsection{Laughlin states}
\label{sub:Laughlin-1_3}
\begin{figure*}[t]
    \centering
    \includegraphics[width=1\linewidth]{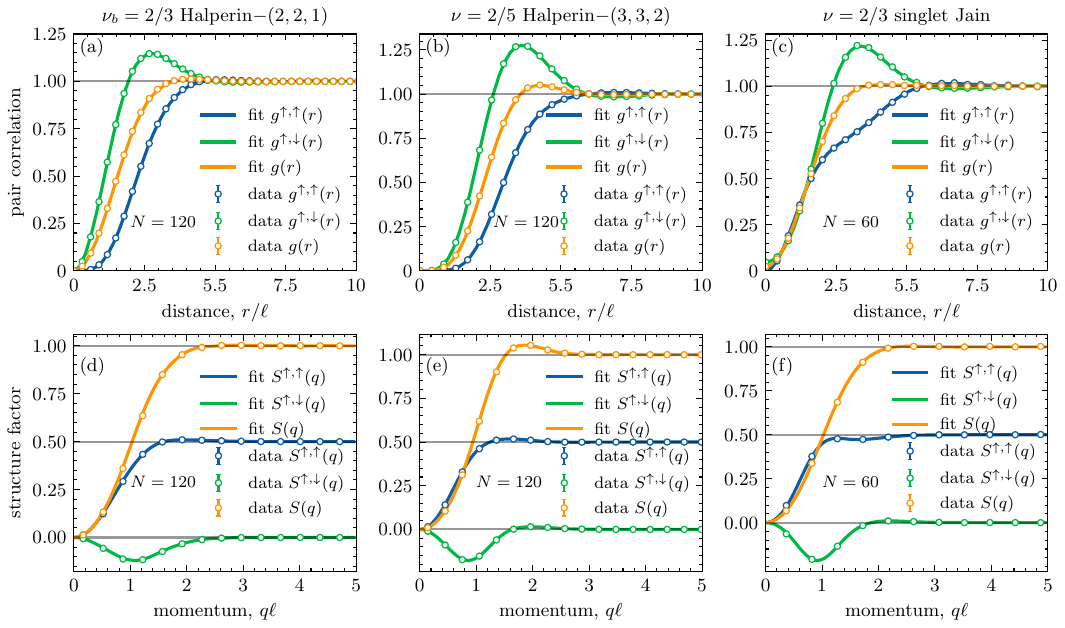}
    \caption{\textbf{Pair-correlation functions and static structure factors of spin-singlet FQH states.}
Panels (a)-(c): The various components of the pair-correlation function for the bosonic $\nu_b{=}2/3$ Halperin-$(2,2,1)$, and the fermionic $\nu{=}2/5$ Halperin-$(3,3,2)$ and the fermionic $\nu{=}2/3$ spin-singlet Jain states, respectively. The circles show the raw Monte Carlo (MC) data obtained on the sphere, while the solid lines are planar fits to these MC data carried out using the expansions in \cref{eq:g-OE,eq:gab-OE}. 
Panels (d)-(f): The corresponding components of the static structure factor for the states shown in panels (a)-(c), respectively. The solid lines are obtained from \cref{eq:S-OE,eq:Sab-OE}, while the circles show the raw MC data obtained on the sphere.
In panels (a)-(f), the error bars are smaller than the plot markers.
}
    \label{fig:fit-result}
\end{figure*}

The numerical results for the $\nu=1/3$ and $1/5$ Laughlin states with $n_{\rm max}=20$ fitting parameters are shown in \cref{fig:Laughlin_fitting}. The MC data against which the fitting is done is for $N=60$ particles on the sphere. We consider two fitting schemes. Fit-1 imposes only the constraint $s_0=0$, \cref{eq:s0}, following Ref.~\cite{Fulsebakke23}, whereas Fit-2 imposes the additional constraints in \cref{eq:s_coeffs_constraints} using the known coefficients $s_2$, $s_4$, and $s_6$ [see \cref{eq:s4-s6-derive,tab:coeffs}] and the correct short-distance behavior, $g_{\nu=1/m}^{\mathrm{Laughlin}}(r\to0)\sim r^{2m}$. For $\nu=1/3$, we use the orthogonal-expansion form in \cref{eq:g-OE} with both schemes, while for $\nu=1/5$ we use Girvin's form in \cref{eq:gcorr} with Fit-2. 

For $\nu=1/3$, both schemes reproduce the MC $g(r)$ with comparable accuracy [\cref{fig:Laughlin_fitting}(a)]. The Coulomb energies are $-0.408\,e^2/\varepsilon\ell$ and $-0.410\,e^2/\varepsilon\ell$ for Fit-1 and Fit-2, respectively, both consistent with the established $E^C$~\cite{Girvin86, Ciftja03, Dora23, Balram20b}. For $\nu=1/5$, Fit-2 gives $E^{C}=-0.327\,e^2/\varepsilon\ell$, in agreement with previous results~\cite{Knoester16, Dora23} [\cref{fig:Laughlin_fitting}(d)].

Using the fitted parameters, we obtain the $S(q)$ from \cref{eq:SF-total,eq:S-OE} and compare it with the MC results [\cref{fig:Laughlin_fitting}(b),(e)]. The two fits give similar results at finite momentum, but differ significantly in the long-wavelength limit. Fit-1 produces the incorrect long wavelength behavior $\bar{S}(q\to0)\sim q^2$, whereas Fit-2 satisfies the required sum rules and recovers $\bar{S}(q\to0)\sim q^4$~\cite{Girvin86}.

This difference results in different GMP gaps, which are calculated using \cref{eq:SDW-gap}, and are shown in \cref{fig:Laughlin_fitting}(c),(f). The oscillator-strength integral in \cref{eq:oscillator-strength} is sensitive to the large-$q$ behavior of $S(q)$, so we regularize it following Ref.~\cite{Fulsebakke23} by restricting the integration to $\abs{\bm{k}+\bm{q}}\leq k_U$. We find stable results for $3.5\leq k_U\leq4.5$. For $\nu=1/3$, Fit-1 gives a vanishing gap as $q\to0$. This follows from $\bar{S}(q)\sim q^2$ and $\bar{F}(q)\sim q^4$, which yield $\Delta^{\mathrm{SDW}}(q)\sim q^2$. In contrast, Fit-2 recovers the correct $\bar{S}(q)\sim q^4$ behavior and gives a finite long-wavelength gap, $\Delta_{\nu=1/3}^{\mathrm{SDW}}(q\to0)\approx0.15\,e^2/\varepsilon\ell$. For $\nu=1/5$, Fit-2 gives $\Delta_{\nu=1/5}^{\mathrm{SDW}}(q\to0)\approx0.05\,e^2/\varepsilon\ell$, consistent with known results~\cite{Girvin86}. The resulting dispersions also agree well with previous Fock-space calculations~\cite{He94, Yang12b, Balram24} and spherical-geometry results~\cite{Dora24, Dora25}. For comparison, the planar momentum $q$ and spherical angular momentum $L$ are related by $q=\sqrt{L(L+1)}/R$~\cite{Anakru25, Dora25}, where $R$ is the radius of the sphere.

At larger momenta, the two fitting schemes give nearly identical dispersions and reproduce the magneto-roton minimum. Thus, the additional sum-rule constraints primarily affect the long-wavelength behavior, while leaving the finite-momentum dispersion largely unchanged.

The different sensitivity of the energy and the GMP gap to these constraints can be understood from their dependence on correlations. For interactions dominated by their short-range part, like Coulomb, the ground-state energy is predominantly determined by short-range correlations and thus, $E^C$ is relatively insensitive to the precise long-wavelength behavior of $S(q)$. In contrast, the GMP gap at small-$q$ depends directly on the small-$q$ behavior of $\bar{S}(q)$, which, aside from being in $\bar{F}(q)$, sits in the denominator of the GMP gap equation [see \cref{eq:SDW-gap}]. Therefore, getting the coefficients $s_2$, $s_4$, and $s_6$ right is essential for obtaining the correct $q{\to}0$ GMP gap.

Finally, we compute the SFDW gaps for the $\nu{=}1/3$ and $1/5$ Laughlin states using \cref{eq:SFDW-gap}, with the structure factors obtained from the fitted parameters [see \cref{fig:fit-result}(b),(e)]. We also show the $\nu=1$ SFDW gap computed using \cref{eq: SFDW_nu_1}. The planar results are compared with sphere calculations using the methodology of \cite{Nakajima94, Dora25} in \cref{fig:SFDW-gaps}. Using \cref{eq:spin-stiffness}, we obtain spin stiffnesses of $0.035\,e^2/\varepsilon\ell$ and $0.015\,e^2/\varepsilon\ell$ for $\nu=1/3$ and $1/5$, respectively. The $\nu=1/3$ value agrees with Ref.~\cite{Zhou22}.

\subsection{Spin-singlet states}
\label{sub:spin-singlet states}

In this subsection, we present the $g^{\alpha, \beta}(r)$ and $S^{\alpha, \beta}(q)$ for several spin-singlet states. For all states, the universal constraints $s_0=0$ and $s_2=1/2$ are imposed during fitting. For chiral states, we additionally impose the $s_4$ and $s_6$ constraints listed in \cref{eq:s4-s6-derive,tab:coeffs}. We also enforce the exact short-distance behavior of $g^{\alpha, \beta}(r)$. For instance, for the Halperin $(m,m,m-1)$ states, $g^{\alpha,\beta}(r)\sim r^{2K_{\alpha,\beta}}$, where $K_{\alpha,\beta}$ is the corresponding element of the $K$ matrix of \cref{eq:Halperin-Kmatrix}.

We first consider the Halperin $(m,m,m-1)$ states for $m=2$ and $m=3$, which correspond to bosons at $\nu_b=2/3$ and fermions at $\nu=2/5$. The MC data for these are generated on the sphere for $N=120$ particles using the wavefunction given in Eq.~\eqref{eq: Halperin_mmpn_wavefunction}. The fitted $g^{\alpha, \beta}(r)$ are shown in \cref{fig:fit-result}(a) and (b). In all cases, the fits are in excellent agreement with the MC data. Using the fitted $g(r)$ in \cref{eq:energy-gr}, we obtain Coulomb energies of $-0.528\,e^2/\varepsilon\ell $ and $-0.438\,e^2/\varepsilon\ell$ for the Halperin-$(2,2,1)$ and $(3,3,2)$ state, respectively. Both values are consistent with thermodynamic estimates obtained from extrapolation of finite-size results~\cite{Balram15a, Dora25}.

We next consider the fermionic Jain spin-singlet state at $\nu=2/3$. Since the wavefunction of Eq.~\eqref{eq: wfn_Jain_2_3_spin_singlet} requires explicit projection, the MC data are generated on the sphere for $N=60$ particles. Unlike the $(m,m,m-1)$ Halperin states, the Jain singlet is non-chiral, so the exact coefficients $s_4$ and $s_6$ for it are not known. Consequently, only the universal constraints $s_0=0$ and $s_2=1/2$ are imposed during the fitting. In addition, no exact long-wavelength constraints are currently available for $S^{\alpha, \beta}(q)$. Another important difference in comparison to the $(m,m,m-1)$ Halperin states is that the $g^{\uparrow, \downarrow}(r)$ for the $2/3$ Jain spin-singlet state remains finite at the origin. The fitted $g^{\alpha, \beta}(r)$ are compared with the MC data in \cref{fig:fit-result}(c), showing excellent agreement despite the fewer available constraints. Using the fitted $g(r)$ in \cref{eq:energy-gr}, we obtain the Coulomb energy of $-0.522\,e^2/\varepsilon\ell$, which is consistent with previous estimates~\cite{Balram15a}. As we discuss in~\cref{sec:results-collective-excitations}, the absence of higher-order long-wavelength constraints on $S^{\alpha, \beta}(q)$ for the $2/3$ Jain spin-singlet state has important repercussions for the calculation of its density-wave excitation gaps.

The lower panels of \cref{fig:fit-result}(d)-(f) show the $S^{\alpha, \beta}(q)$ obtained from the $g^{\alpha, \beta}(r)$ in \cref{fig:fit-result}(a)-(c). They are obtained from \cref{eq:s-OE} using the same fitting coefficients $\{d_n\}$ and $\{d_n^{\alpha,\beta}\}$ extracted from the $g^{\alpha, \beta}(r)$ fits. As expected, $S(q)\to1$ as $q\to\infty$, while $S^{\uparrow,\uparrow}(q)\to1/2$ and $S^{\uparrow,\downarrow}(q)\to0$. In all cases, the fitted $S^{\alpha, \beta}(q)$ are in excellent agreement with the MC data over all wavenumbers.
\begin{table}[htpb]
\centering
\begin{tabular}{c c c c c c}
\toprule
{state} & {component} & $s^{\alpha,\beta}_0$ & $s^{\alpha,\beta}_2$ & $s^{\alpha,\beta}_4$ & $s^{\alpha,\beta}_6$ \\
\midrule
$\nu=1/3$ & total & $0$ & $1/2$ & $1/8$ & $0$ \\
Laughlin  & &  & & & \\
\midrule
$\nu=1/5$ & total & $0$ & $1/2$ & $3/8$ & $11/48$ \\
Laughlin  & &  & & & \\
\midrule
$\nu_b=2/3$  & total & $0$ & $1/2$ & $0$ & $1/32$ \\
Halperin-$(2,2,1)$ & $\up,\up$ & $0$ & $1/2$ & $-$ & $-$ \\
               & $\up,\dn$ & $0$ & $-1/4$ & $-$ & $-$ \\
\midrule
$\nu=2/5$ & total & $0$ & $1/2$ & $1/8$ & $-1/48$ \\
Halperin-$(3,3,2)$ & $\up,\up$ & $0$ & $3/4$ & $-$ & $-$ \\
 & $\up,\dn$ & $0$ & $-1/2$ & $-$ & $-$ \\
\midrule
$\nu=2/3$ & total & $0$ & $1/2$ & $-$ & $-$ \\
Jain singlet  & $\up,\up$ & $0$ & $-$ & $-$ & $-$ \\
               & $\up,\dn$ & $0$ & $-$ & $-$ & $-$ \\
\bottomrule
\end{tabular}
\caption{Exact leading long-wavelength coefficients of the static structure factor used to constrain the fits to the Monte Carlo pair-correlations. Entries marked by ``$-$'' denote coefficients for which no exact results are available.}
\label{tab:coeffs}
\end{table}

\begin{figure*}[htpb]
    \centering
    \includegraphics[width=\linewidth]{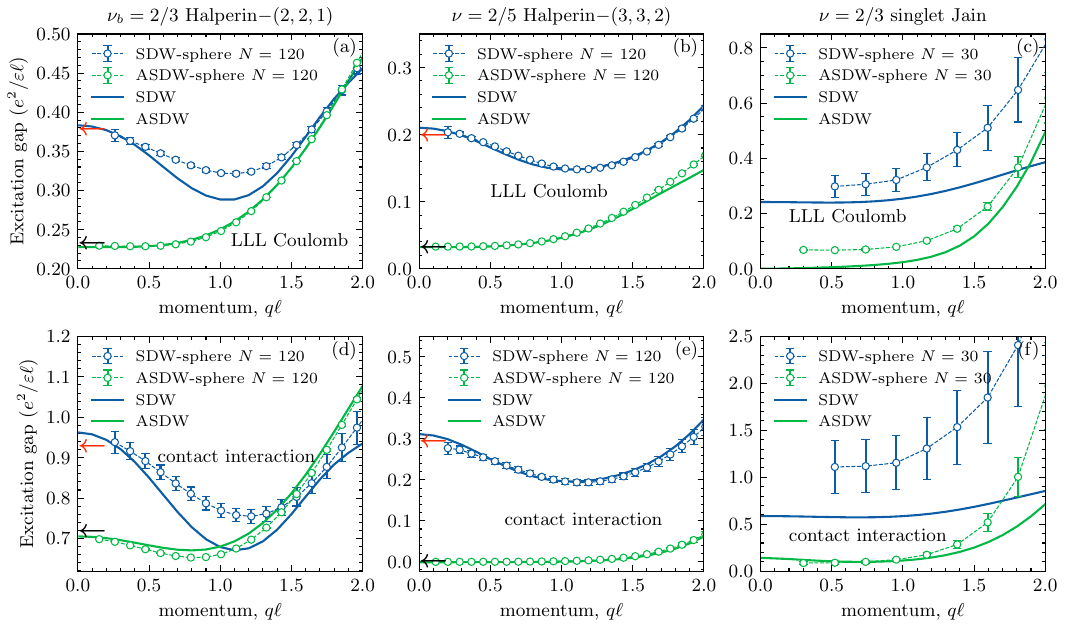}
    \caption{
    \textbf{Density-wave dispersions for spin-singlet FQH states.}
Panels (a)-(c): Dispersions of the symmetric-density-wave (SDW) and anti-symmetric-density-wave (ASDW) modes for the bosonic $\nu_b{=}2/3$ Halperin-$(2,2,1)$, and the fermionic $\nu{=}2/5$ Halperin-$(3,3,2)$ and the fermionic $\nu{=}2/3$ spin-singlet Jain states, respectively computed using \cref{eq:SDW-gap,eq:ASDW-gap} for the Coulomb interaction. We also show the corresponding gaps on the sphere. The error bar on the spherical gaps stems from the statistical uncertainty in the Monte Carlo evaluation of the structure factor. The spherical gaps are computed using methods outlined in Ref.~\cite{Dora25}. 
Panels (d)-(f) present the corresponding results for the contact interaction. The planar ASDW gap for the $\nu{=}2/3$ spin-singlet Jain state is erroneous because the sum rules are not known for this state. The red and black arrows in (a), (b), (d), and (e) mark $q{\to}0$ gaps computed using \cref{eq:k1_SDW_gap,eq:k0_ASDW_gap}, respectively.}
    \label{fig:gaps}
\end{figure*}

\section{Results: Density-wave modes}
\label{sec:results-collective-excitations}

In this section, we present the computed dispersions of the density-wave modes for both the Coulomb and contact interactions and compare them with the corresponding results obtained on the sphere~\cite{Dora25}. The SDW and ASDW gaps are computed using \cref{eq:SDW-gap,eq:ASDW-gap}, employing the same regularization scheme discussed in \cref{sub:Laughlin-1_3}. Specifically, the oscillator-strength integral is restricted to $\abs{\bm{k}+\bm{q}}\leq k_U$. The cutoff $k_U$ is chosen such that the long-wavelength SDW and ASDW gaps agree with the thermodynamic extrapolations obtained on the sphere, shown in \cref{fig:thermodynamic-gap} and in Ref.~\cite{Dora25}, following the methodology therein. 

On the sphere, the long-wavelength SDW and ASDW gaps are extracted from the lowest allowed angular-momentum of the excitation, namely $L=2$ and $L=1$, respectively~\cite{Dora25}. Accessing the $q\to0$ [recall that $q=\sqrt{L(L+1)}/R$~\cite{Anakru25, Dora25}, where $R$ is the sphere's radius] regime requires going to prohibitively large systems on the sphere. In contrast, the fitted $S^{\alpha,\beta}(q)$ described in \cref{sub:spin-singlet states} provides a natural way to access this long-wavelength regime on the plane.

The dispersion of the density-wave modes for the bosonic $\nu_b=2/3$ Halperin-$(2,2,1)$ state is shown in \cref{fig:gaps}(a) and (d) for the Coulomb and contact interactions, respectively. The ASDW dispersion is in excellent agreement with the gaps obtained from the sphere over the entire momentum range. In contrast, although the planar SDW correctly captures the long- and short-wavelength limits, the largest deviation from the spherical results occurs near the magnetoroton minimum. We attribute this discrepancy to finite-size curvature effects on the sphere.

The dispersion of the density-wave modes for fermionic $\nu=2/5$ Halperin-$(3,3,2)$ state is shown in \cref{fig:gaps}(b) and (e) for the Coulomb and contact interactions, respectively. Among the states considered here, this system is studied on the largest sphere. As a result, finite-curvature effects are substantially reduced, and both the planar SDW and ASDW dispersions are in excellent agreement with the spherical ones over the entire momentum range. For the contact interaction, we further find that the ASDW mode becomes gapless in the long-wavelength limit. This can be understood by noting that the ASDW mode and the corresponding anti-symmetric composite fermion exciton for the Halperin-$(2,2,1)$ and Halperin-$(3,3,2)$ states are identical in the long-wavelength limit~\cite{Dora25} [but are not in agreement with each other at finite-$q$]. Attaching a vortex to the ASDW state of Halperin-$(2,2,1)$ results in the following hard-core projected~\cite{Wu93} ansatz for the long-wavelength ASDW mode for the Halperin-$(3,3,2)$ state
\begin{align}
    \lim_{q\to0}\Psi_{2/5}^{\mathrm{ASDW}}(q)
    = \lim_{q\to0}\bar{\rho}^z_q \Psi^{(3,3,2)}_{\nu=2/5}
    = \Phi_1\lim_{q\to0} \mathcal{P}_{\mathrm{LLL}}  \bar{\rho}^z_q
    \Psi^{(2,2,1)}_{\nu_b=2/3}.
    \label{eq: ASDW_qto0_ansatz}
\end{align}
The Jastrow factor $\Phi_1$ ensures that the wavefunction of Eq.~\eqref{eq: ASDW_qto0_ansatz} vanishes when any two particles, irrespective of their spins, coincide. Consequently, its energy expectation value for the contact interaction vanishes as seen in \cref{fig:gaps}(e). It is important to note that the SDW mode is not the lowest-lying neutral excitation at any $q$ in the Halperin-$(3,3,2)$ state, and the SDW splits into two modes, rendering the SDW-SMA invalid for the Halperin-$(3,3,2)$ state~\cite{Dora25}. Such a splitting does not happen for the ASDW mode, and essentially we just appealed to this fact to construct the wavefunction given in Eq.~\eqref{eq: ASDW_qto0_ansatz}.

Finally, the excitation spectra for the fermionic $\nu=2/3$ Jain spin-singlet state are shown in \cref{fig:gaps}(c) and (f) for the Coulomb and contact interactions, respectively. Here, the agreement between the planar and spherical results is noticeably poorer. This can be attributed to two factors. First, unlike the chiral Halperin spin-singlet states, no exact sum rules are currently known for the long-wavelength coefficients $s_4$ and $s_6$ of the $2/3$ Jain singlet state (see \cref{tab:coeffs}), reducing the accuracy of the fitted $S^{\alpha, \beta}(q)$ at small momenta. Second, very accurate numerical data is available only for relatively small systems ($N=30$) [Note that in~\cref{fig:fit-result}(c) and (f), we show results for $N=60$, but those numbers are not accurate enough to get reliable density-wave gaps.], where finite-curvature effects remain substantial. Most notably, the ASDW mode should be gapped for the Coulomb interaction, as seen in the spherical results, but the planar ASDW mode has a vanishing gap as $q{\to}0$. Thus, we conclude that the planar gaps shown in \cref{fig:gaps}(c) and (f) are unreliable. 

For all the spin-singlet states considered in this work, the ASDW mode has lower energy than the SDW at small momenta, and in particular in the long-wavelength limit, as evidenced by the results shown in \cref{fig:gaps} and~\cref{fig:thermodynamic-gap} and previous results of Ref.~\cite{Dora25}. For the Halperin-$(m,m,m-1)$ states, this ordering is also predicted by the effective field theory discussed in \cref{sub:application-to-Halperin}.

\begin{figure*}[htpb]
    \centering
    \includegraphics[width=1\linewidth]{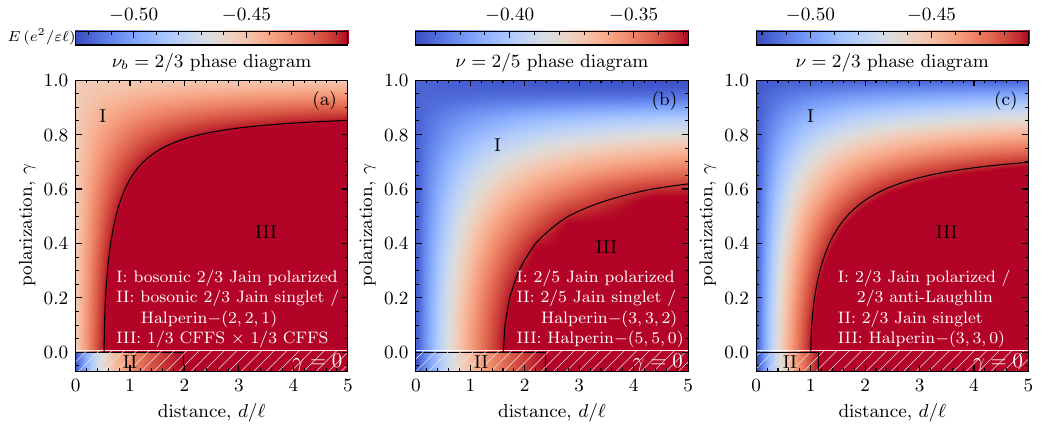}
\caption{
\textbf{Variational phase diagrams of bilayer FQH states.}
Phase diagrams in the interlayer separation-polarization \((d/\ell,\gamma)\) plane for (a) bosons at filling \(\nu_b{=}2/3\), and fermions at fillings (b) \(\nu{=}2/5\) and (c) \(\nu{=}2/3\) obtained from calculations comparing the per-particle Coulomb energies of the competing candidate states listed on the panels and identifying the state with the lowest energy at each point. Solid lines denote phase boundaries. At non-zero polarizations (\(\gamma{>}0\)), only the fully polarized (I) and layer-decoupled (III) states compete, whereas at \(\gamma{=}0\) the singlet (II) state is also relevant. The hatched area shown below \(\gamma{=}0\) highlights the phase diagram on the \(\gamma{=}0\) line.}
    \label{fig:phase-diagram}
\end{figure*}

\section{Variational phase diagram of ground state}
\label{sec:phase-diagram}

In this section, we turn to bilayer FQH systems, where the layer degree of freedom serves as a pseudospin, and the real spin is assumed to be fully polarized. We identify $\uparrow$ ($\downarrow$) with the top (bottom) layer, and denote the corresponding particle number by $N_{\uparrow}$ ($N_{\downarrow}$). We assume the absence of interlayer tunneling, which can be suppressed by adding a hBN spacer layer in graphene double layers, as has been done in the recent experiments reported in Refs.~\cite{Nguyen24a, Nguyen26}. Two parameters that characterize the system are the layer polarization $\gamma=(N_{\uparrow}-N_{\downarrow})/(N_{\uparrow}+N_{\downarrow})$ and the interlayer separation $d$ [$d/\ell$, but we set $\ell=1$]. Our goal is to determine the variational phase diagram by comparing the energies of several candidate wavefunctions using \cref{eq:energy-gr}. The intralayer and interlayer Coulomb interactions are
$v^{\uparrow,\uparrow}(r)=1/r=v^{\downarrow,\downarrow}(r)$
and
$v^{\uparrow,\downarrow}(r)=1/\sqrt{r^2+d^2}=v^{\downarrow,\uparrow}(r)$,
respectively. Since these interactions differ, the Hamiltonian breaks $SU(2)$ pseudospin symmetry, although this symmetry is approximately restored for $d\ll\ell$. The phase diagrams obtained below should therefore be regarded as variational since the exact ground states need not have good pseudospin quantum numbers, while the singlet candidates do.

The resulting phase diagrams are shown in \cref{fig:phase-diagram}. The hatched region below $\gamma=0$ is the phase diagram along the $\gamma=0$ axis. For each filling, we compare three classes of trial states: (I) fully pseudospin-polarized states, (II) pseudospin-singlet states, and (III) layer-decoupled states. The latter generally require different magnetic fields for the two layers when $\gamma\neq0$, since the layers contain different numbers of particles while remaining at the same filling. At zero polarization, the singlet states compete directly with the decoupled states, whereas for finite polarization, the polarized states compete with the layer-decoupled states. As $d$ increases, the decoupled phase eventually becomes energetically favorable. The energies of all the fully polarized states at arbitrary $\gamma$ are obtained from the $g(r)$ of the fully polarized state ($\gamma=1$) using the relations derived in Ref.~\cite{Kundu26}.

Several features can be gleaned from these phase diagrams. At $d=0$, the $SU(2)$ pseudospin symmetry is restored, making the energy independent of $\gamma$. Conversely, at $\gamma=1$, all particles occupy a single layer, so the interlayer interaction plays no role and the energy becomes independent of $d$. For intermediate $\gamma$ and sufficiently large $d$, the decoupled phase is always favored.

For the bosonic $\nu_b=2/3$ system, we compare three candidate states: (I) the fully polarized $\nu_b=2/3$ Jain state, (II) the $\nu_b=2/3$ Jain spin-singlet state, which is equivalent to the Halperin $(2,2,1)$ state, and (III) the product of two decoupled bosonic composite-fermion Fermi-sea (CFFS) state, each at $\nu_b=1/3$~\cite{Balram17}, denoted by $1/3\,\mathrm{CFFS}\times1/3\,\mathrm{CFFS}$, which has $E^C=-0.405\,e^2/\varepsilon\ell$. The resulting phase diagram is shown in \cref{fig:phase-diagram}(a). Since the $g(r)$ of the polarized Jain state remains finite at $r=0$, whereas that of the singlet state vanishes, the energy difference between these phases is relatively large near $\gamma=0$. At $\gamma=0$, the singlet state is favored for small $d$, while a transition to the decoupled CFFS state occurs at $d/\ell\approx2$.

For the fermionic $\nu=2/5$ system, the competing phases are (I) the fully polarized $\nu=2/5$ Jain state, (II) the $\nu=2/5$ Jain spin-singlet state, equivalent to the Halperin-$(3,3,2)$ state, and (III) the decoupled Halperin-$(5,5,0)$ state, which has $E^C=-0.328\,e^2/\varepsilon\ell$ [As explained in Ref.~\cite{Kundu26}, for layer-decoupled states, $E^C$ is independent of $d$ and $\gamma$.]. The corresponding phase diagram is shown in \cref{fig:phase-diagram}(b). Unlike the bosonic case, both the polarized and singlet states have $g(r)$ that vanishes at $r=0$, resulting in a much smaller energy difference between them. At $\gamma=0$, the singlet phase remains stable up to $d/\ell\approx2.4$, beyond which the decoupled Halperin-$(5,5,0)$ state prevails.

For the fermionic $\nu=2/3$ system, we compare (I) the fully polarized $\nu=2/3$ Jain state, which is nearly identical to the hole-conjugate of the $\nu=1/3$ Laughlin state~\cite{Balram21b}, termed $2/3$ anti-Laughlin state, (II) the $\nu=2/3$ Jain spin-singlet state, and (III) the decoupled Halperin-$(3,3,0)$, which has $E^C=-0.410\,e^2/\varepsilon\ell$~\cite{Girvin86, Ciftja03, Dora23, Balram20b}. The resulting phase diagram is shown in \cref{fig:phase-diagram}(c). At $\gamma=0$, the singlet state is favored for small $d$, which transitions to the Halperin $(3,3,0)$ state at $d/\ell\approx1.2$. We note that our results are in agreement with previous calculations that were also carried out in the zero interlayer tunneling limit, but used finite-size results and then extrapolated those to the thermodynamic limit~\cite{Peterson15, Geraedts15, Faugno20}.

\section{Conclusions}
\label{sec:conclusion}

In this work, we employed the orthogonal basis of Fulsebakke~\emph{et al.}~\cite{Fulsebakke23} to obtain an accurate parameterization of the pair-correlation function $g(r)$ of chiral fractional quantum Hall (FQH) states by enforcing the exact long-wavelength constraints that its Fourier transform, the static structure factor $S(q)$, must adhere to. Capturing the correct long-wavelength behavior of $S(q)$ is crucial for accurately determining the neutral density-wave excitation gap in the long-wavelength limit.

Using effective field theory, we derived the small-$q$ expansion of the $S(q)$. We extended the orthogonal basis to the spin-resolved pair-correlation functions $g^{\alpha,\beta}(r)$ ($\alpha, \beta \in [\uparrow, \downarrow$]) by imposing the corresponding constraints on the spin-resolved structure factors $S^{\alpha,\beta}(q)$. This allowed us to calculate the symmetric and antisymmetric density-wave (SDW and ASDW) excitation spectra for several candidate spin-singlet FQH states. The long-wavelength gaps obtained here can be directly compared against inelastic light scattering or surface-acoustic wave measurements that probe the collective modes~\cite{Pinczuk93, Kukushkin09, Liang24, Yang26}, which may help distinguish between competing candidate states and provide experimental evidence for the stabilization of particular FQH phases.

The accurate parameterizations of $g^{\alpha,\beta}(r)$ also allow reliable calculations of ground-state energies for arbitrary two-body interaction potentials. This is particularly useful for the bilayer setting, where the interaction depends on the layer separation $d$, and the layer polarization can be controlled by tuning the density imbalance via electrostatic gates. By comparing the energies of competing candidate states across this parameter space, we constructed variational phase diagrams that identify the phases most likely to be stabilized in bilayer FQH systems at selected fillings. 

In Appendix~\ref{app: density_wave_gaps_SU2_breaking}, we have extended the density-wave gap computations to Hamiltonians that break $SU(2)$ symmetry, such as in the bilayer setting. Another important direction for future work is the derivation of exact higher-order sum rules for spin-resolved density-density correlators. Such relations would further constrain the density-density correlation functions and could substantially improve the quantitative description of density-wave modes. More generally, the parametrization developed here can be extended to multicomponent FQH systems with additional internal degrees of freedom, such as valley and orbital pseudospin, that arise in systems based on multilayer graphene~\cite{Xie21, Lu24, Dong24, Xie25, Lu25,  Li26} and transition metal dichalcogenides~\cite{Cai23, Zeng23, Xu23, Park23}. Recently, zero-field analogs of FQH states have been realized in these systems, but computing density-wave modes in these settings remains numerically challenging~\cite{Kousa25, Long26, Goncalves26, Shen26}. Our approach, therefore, provides a unified framework for studying ground-state density-density correlations and neutral density-wave modes dispersions in these and other systems.

\section*{Data availability}
All the data used in this study are
publicly available via Zenodo at \url{https://doi.org/10.5281/zenodo.21885623}.

\begin{acknowledgments}
We acknowledge valuable discussions with G. J. Sreejith, Jainendra Jain, Koyena Bose, and Prashant Kumar. The work was made possible by financial support from the Anusandhan National Research Foundation (ANRF) of the Department of Science and Technology (DST) via the Mathematical Research Impact Centric Support (MATRICS) Grant No. MTR/2023/000002 and the Advanced Research Grant No. ANRF/ARG/2025/000562/PS and National Postdoctoral Fellowship Grant No. PDF/2025/002176. This research was supported in part by the International Center for Theoretical Sciences (ICTS) through DXN's and ACB's participation in the program ``Generalized symmetries and anomalies in quantum phases of matter 2026" (code: ICTS/GSYQM2026/01). Computational portions of this research work were conducted using the Nandadevi and Kamet supercomputers maintained and supported by the Institute of Mathematical Sciences' High-Performance Computing Center. ACB is grateful to the Lodha Theoretical Physics Institute (LTPI) for its hospitality, where a part of this work was completed. 
\end{acknowledgments}

\appendix
\crefalias{section}{appendix}
\setcounter{figure}{0}
\renewcommand{\thefigure}{S\arabic{figure}}
\section{Relation between the projected and unprojected structure factors}
\label{app:Relation between the projected and unprojected structure factors}
In this appendix, we derive the relations between the projected and unprojected structure factors used in the main text. We begin by considering the operator
\begin{align}
    \rho^{\alpha}_{-q}\rho^{\beta}_{q}
    =
    \sum_{i = 1}^{N} \sum_{j=1}^N
    \sigma_i^\alpha \sigma_j^\beta
    \otimes
    e^{-\frac{\iota}{2}(q z_i^* + q^* z_i)}
    e^{\frac{\iota}{2}(q z_j^* + q^* z_j)},
    \label{eq: product_densities}
\end{align}
where \(\rho_q^\alpha\) is the spinful density operator introduced in the main text [see Eq.~\eqref{eq:density-operator}]. The indices \(\alpha,\beta {\in} \{I,x,y,z,\up,\dn\}\), with \(\sigma^\up = (\sigma^I{+}\sigma^z)/2\) and \(\sigma^\dn{=}(\sigma^I{-}\sigma^z)/2\).

Projecting the operator in Eq.~\eqref{eq: product_densities} onto the LLL using the Girvin-Jach prescription~\cite{Girvin84b} and rewriting the result in terms of magnetic translation operators yields
\begin{align}
    \overline{\rho^\alpha_{-q}\rho^\beta_q}
    =
    e^{-q^2/2}
    \sum_{ij}
    \sigma_i^\alpha \sigma_j^\beta
    \otimes
    \tau_{-q}(i)\tau_q(j)
    e^{\frac{q^2}{2}\delta_{ij}},
\end{align}
where \(\overline O {\equiv} \mathcal P_{\mathrm{LLL}} O \mathcal P_{\mathrm{LLL}}\) denotes the LLL projection of an operator $O$.

Using the definition of the projected density operator [see Eq.~\eqref{eq:projected-density-operator}], this expression can be rewritten as
\begin{align}
    \overline{\rho^\alpha_{-q}\rho^\beta_q} = \rh{\alpha}{-q}\rh{\beta}{q} + S_{\nu=1}(q) \sum_i \sigma_i^\alpha \sigma_i^\beta, \label{eq:projected-unprojected-density}
\end{align}
where \(S_{\nu=1}(q){=}1{-}e^{{-}q^2/2}\) is the structure factor of the \(\nu{=}1\) filled LLL. Eq.~\eqref{eq:projected-unprojected-density} illustrates an important point: the LLL projection of a product of operators is generally not equal to the product of the individually projected operators~\cite{Girvin86, Dora24}.

For a state \(\ket{\Psi}\) that resides entirely within the LLL, \(\mathcal P_{\mathrm{LLL}}\ket{\Psi}=\ket{\Psi}\). Therefore,
\begin{align}
    \expval{\rho^\alpha_{-q}\rho^\beta_q}{\Psi}
    &=
    \expval{ \mathcal P_{\mathrm{LLL}} \rho^\alpha_{-q}\rho^\beta_q \mathcal P_{\mathrm{LLL}} }{\Psi}
    \nonumber \\
    &= \expval{\overline{\rho^\alpha_{-q}\rho^\beta_q}}{\Psi}
    \\
    &= \expval{\rh{\alpha}{-q}\rh{\beta}{q}}{\Psi} + S_{\nu=1}(q) \expval{\sum_i \sigma_i^\alpha \sigma_i^\beta}{\Psi}. \nonumber
\end{align}

The unprojected structure factor is defined as
\begin{align}
    S^{\alpha,\beta}(q) = \frac{1}{N} \expval{\rho^\alpha_{-q}\rho^\beta_q}{\Psi}.
\end{align}
It follows immediately that
\begin{align}
    S^{\alpha,\beta}(q) = \bar S^{\alpha,\beta}(q) + S_{\nu=1}(q) \frac{ \expval{\sum_i \sigma_i^\alpha \sigma_i^\beta}{\Psi} }{N},
\end{align}
where
\(
\bar S^{\alpha,\beta}(q) = \expval{\bar\rho^\alpha_{-q}\bar\rho^\beta_q}{\Psi}/N
\)
is the projected structure factor.

The relations used in the main text are obtained as special cases~\cite{Kundu26}:
\begin{align}
    S^{\up,\up}(q) &= \bar S^{\up,\up}(q) + \frac{N_\up}{N} S_{\nu=1}(q),
    \\
    S^{\dn,\dn}(q) &= \bar S^{\dn,\dn}(q) + \frac{N_\dn}{N} S_{\nu=1}(q),
    \\
    S^{\up,\dn}(q) &= \bar S^{\up,\dn}(q) = \bar S^{\dn,\up}(q),
\end{align}
where \(N_\up\) (\(N_\dn\)) denotes the number of spin-\(\up\) (spin-\(\dn\)) particles in \(\ket{\Psi}\). Similarly,
\begin{align}
    S^{I}(q) &= \bar S^{I}(q) + S_{\nu=1}(q),
    \\
    S^{z}(q) &= \bar S^{z}(q) + S_{\nu=1}(q),
\end{align}
where \(S^{I}(q){=}\expval{\rho^I_{-q}\rho^I_q}{\Psi}/N\) and
\(S^{z}(q){=}\expval{\rho^z_{-q}\rho^z_q}{\Psi}/N\), with analogous definitions for their projected counterparts. Analogous relations for the finite spherical geometry are provided in Refs.~\cite{Dora25, Kundu26}.

\section{Structure factor of multi-layer quantum Hall states from effective field theory}
\label{app:field-theory}
\subsection{Derivation of the gauge-field propagator
\texorpdfstring{$\langle a a \rangle$}{<aa>}}
\label{app:gauge-propagator}

In this appendix, we work out the expansion of the structure factor in the long-wavelength limit ($q\to0$) from effective field theory. To do so, we derive the gauge-field propagators by inverting the quadratic kernel in Eq.~\eqref{eq:kernel}. The propagator is given by
\begin{equation}
G=\iota \,\mathcal K^{-1},
\end{equation}
where $\mathcal K$ has the block form
\begin{equation}
\mathcal K=
\begin{pmatrix}
A & B\\
B^\dagger & D
\end{pmatrix},
\end{equation}
with
\begin{equation}
A=\varepsilon_e q^2,\qquad
B=-\frac{\iota q}{2\pi}K,\qquad
D=\varepsilon_e\omega^2-\mu^{-1}q^2,
\label{eq:blocks}
\end{equation}
where $K$ is the $K$-matrix that describes the topological features of the FQH state, and $\varepsilon_e$ and $\mu^{-1}$ encode its non-universal properties. Using the standard block-matrix inversion formula,
\begin{align}
\mathcal K^{-1}
&=
\begin{pmatrix}
\left(A-BD^{-1}B^\dagger\right)^{-1}
&
-A^{-1}B\,\Delta^{-1}
\\[2mm]
-\Delta^{-1}B^\dagger A^{-1}
&
\Delta^{-1}
\end{pmatrix},
\nonumber\\
\Delta
&\equiv
D-B^\dagger A^{-1}B,
\label{eq:inverse}
\end{align}
and the Schur complement appearing in the transverse sector is
\begin{equation}
B^\dagger A^{-1}B
=
\left(-\frac{\iota q}{2\pi}K\right)
\frac{1}{\varepsilon_e q^2}
\left(\frac{\iota q}{2\pi}K\right)
=
\frac{1}{(2\pi)^2}
K\varepsilon_e^{-1}K.
\label{eq:schur}
\end{equation}

The transverse gauge-field propagator, therefore, takes the form
\begin{equation}
\langle
a_T^{\mathsf I}
a_T^{\mathsf J}
\rangle(\omega,q)
=
\iota 
\left[
\varepsilon_e\omega^2
-\mu^{-1}q^2
-\frac{1}{(2\pi)^2}
K\varepsilon_e^{-1}K
\right]^{-1}_{\mathsf I\mathsf J}.
\end{equation}

\subsection{Density-density correlator and frequency integration}
\label{app:frequency-integral}

To evaluate the frequency integral in \cref{eq:Sdef} to get the static density-density correlator from the dynamic one, we use the identity
\begin{equation}
\int\frac{\dd\omega}{2\pi}\,
\iota \left[
\omega^2-\tilde C+\iota \varepsilon
\right]^{-1}
=
\frac12\,\tilde C^{-1/2},
\label{eq:lemma}
\end{equation}
which holds for any symmetric positive-definite matrix $\tilde C$. Writing
$
\varepsilon_e\omega^2-C
=
\varepsilon_e^{1/2}
\left(
\omega^2-\tilde C
\right)
\varepsilon_e^{1/2},
\qquad
\tilde C
=
\varepsilon_e^{-1/2}
C
\varepsilon_e^{-1/2},
$
we obtain
$
\left[
\varepsilon_e\omega^2-C
\right]^{-1}
=
\varepsilon_e^{-1/2}
\left[
\omega^2-\tilde C
\right]^{-1}
\varepsilon_e^{-1/2}.
$ Applying Eq.~\eqref{eq:lemma} immediately yields the exact expression for the static structure factor given in Eq.~\eqref{eq:Sexact}, with $C$ given in Eq.~\eqref{eq:Cq}.

To extract the long-wavelength limit, we expand
\begin{equation}
C
=
C_0+\mathcal O(q^2),
\qquad
C_0
=
\frac{1}{(2\pi)^2}
K\varepsilon_e^{-1}K,
\end{equation}
where the $\mu^{-1}q^2$ contribution is suppressed by a relative factor of $\mathcal O(q^2)$. Introducing the symmetric matrix
$
\mathcal A
=
\varepsilon_e^{-1/2}
K
\varepsilon_e^{-1/2},
$
one finds
\begin{align}
\varepsilon_e^{-1/2}
C_0
\varepsilon_e^{-1/2}
&=
\frac{1}{(2\pi)^2} \mathcal A^2, \nonumber\\ \Rightarrow \left( \varepsilon_e^{-1/2} C_0 \varepsilon_e^{-1/2} \right)^{-1/2}
&=
2\pi \mathcal A^{-1} = 2\pi \varepsilon_e^{1/2} K^{-1} \varepsilon_e^{1/2}. \label{eq:expand}
\end{align}

The permittivity matrices cancel according to
$
\varepsilon_e^{-1/2}
\mathcal A^{-1}
\varepsilon_e^{-1/2}
=
K^{-1},
$
leading to the long-wavelength static structure factor,
\begin{equation}
S^{\mathsf I,\mathsf J}(q)
=
\frac{q^2 \mathbf{t}_{\mathsf I} \mathbf{t}_{\mathsf J}}{4\pi\rho_0}
(K^{-1})_{\mathsf I\mathsf J}
+\mathcal O(q^4),
\end{equation}
where $\rho_0$ is the average density of the FQH state.

\begin{figure}[htpb]
\centering
\begin{minipage}{0.49\linewidth}
\centering
\begin{overpic}[width=1.1\linewidth]{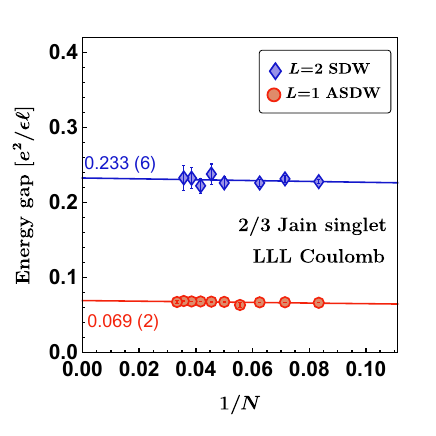}
    \put(20,85){(a)}
\end{overpic}
\end{minipage}
\hfill
\begin{minipage}{0.49\linewidth}
\centering
\begin{overpic}[width=1.1\linewidth]{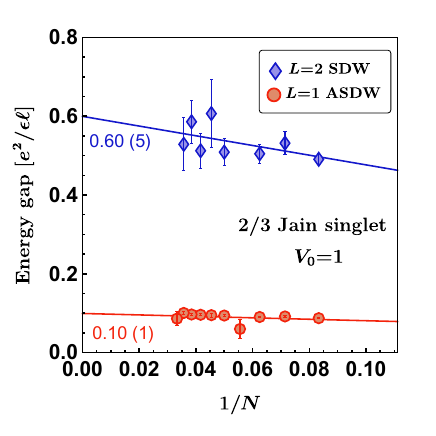}
    \put(20,85){(b)}
\end{overpic}
\end{minipage}

\vspace{-1em}

\begin{minipage}{0.49\linewidth}
\centering
\begin{overpic}[width=1.1\linewidth]{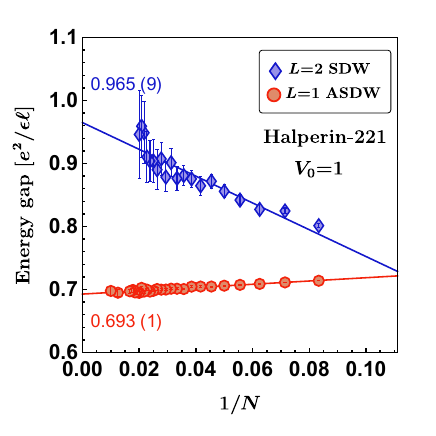}
    \put(20,85){(c)}
\end{overpic}
\end{minipage}
\hfill
\begin{minipage}{0.49\linewidth}
\centering
\begin{overpic}[width=1.1\linewidth]{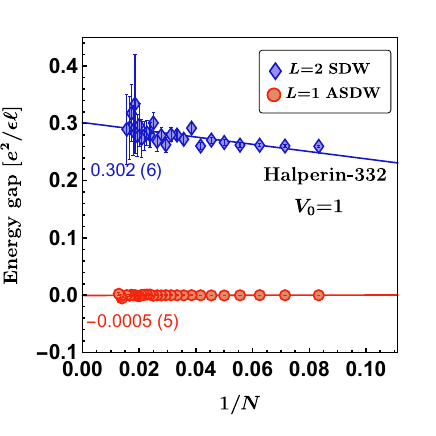}
    \put(20,85){(d)}
\end{overpic}
\end{minipage}

\caption{
\textbf{Thermodynamic extrapolation of the density-corrected neutral density-wave excitation gaps on the sphere.} (a) $L=1$ antisymmetric density-wave (ASDW) and $L=2$ symmetric density-wave (SDW) excitation gaps of the $\nu=2/3$ Jain spin-singlet state for the Coulomb interaction. (b) Corresponding excitation gaps for the $\nu=2/3$ Jain spin-singlet state with the contact interaction, equivalently the Haldane pseudopotential $V_{\mathfrak{m}}=\delta_{\mathfrak{m},0}$. (c) Excitation gaps of the bosonic $\nu_b=2/3$ Halperin $(2,2,1)$ state for the contact interaction. (d) Excitation gaps of the fermionic $\nu=2/5$ Halperin $(3,3,2)$ state for the contact interaction.
}
\label{fig:thermodynamic-gap}
\end{figure}

\section{Thermodynamic extrapolation of density wave gaps}
\label{app:thermodynamic-extrapolation}

In this appendix, we present the thermodynamic extrapolation of the density-corrected~\cite{Morf86b} density-wave gaps obtained in the spherical geometry~\cite{Haldane83} that can be used to benchmark the planar results presented in the main text. The gaps on the sphere are computed and extrapolated using the procedure developed and outlined in Ref.~\cite{Dora25}.

Figure~\ref{fig:thermodynamic-gap}(a) shows the thermodynamic extrapolation of the $L=1$ antisymmetric density-wave (ASDW) gap and the $L=2$ symmetric density-wave (SDW) gap for the $\nu=2/3$ Jain spin-singlet state for the Coulomb interaction. The corresponding results for the contact interaction, $V_{\mathfrak m}=\delta_{\mathfrak m,0}$, are shown in Fig.~\ref{fig:thermodynamic-gap}(b). Figures~\ref{fig:thermodynamic-gap}(c) and (d) display the corresponding extrapolations for the bosonic Halperin-$(2,2,1)$ state at $\nu_b=2/3$ and the fermionic Halperin-$(3,3,2)$ state at $\nu=2/5$, respectively, both for the contact interaction. The thermodynamic extrapolations for the Halperin states with the Coulomb interaction are presented in Ref.~\cite{Dora25}. These extrapolated long-wavelength SDW and ASDW gaps are used throughout the main text to determine the cutoff $k_U$ in Eqs.~\eqref{eq:SDW-gap} and \eqref{eq:ASDW-gap}. Specifically, $k_U$ is chosen such that the long-wavelength gaps obtained from the gap equations reproduce the thermodynamic values shown in Fig.~\ref{fig:thermodynamic-gap}.

\section{Density wave gaps of \texorpdfstring{$SU(2)$}{SU(2)} broken Hamiltonian}
\label{app: density_wave_gaps_SU2_breaking}

In this appendix, we calculate the density-wave gaps for an $SU(2)$-symmetry-violating interaction, such as that arising in the bilayer scenario. Consider the projected interaction Hamiltonian
\begin{align}
    \bar H
    =
    \frac{1}{2}
    \sum_{\alpha,\beta\in\{\up,\dn\}}
    \int\frac{ \dd^2\bm q}{(2\pi)^2}\,
    v^{\alpha,\beta}(q)
    \left[
        \bar\rho^\alpha_{- q}\bar\rho^\beta_{ q}
        -
        \delta_{\alpha,\beta}e^{-q^2/2} \frac{N}{2}
    \right].
    \label{eq:projected-Hamiltonian-general}
\end{align}
We assume the two pseudospins to be in the same setting, but the interactions break $SU(2)$ symmetry, so that
\begin{align}
    v^{\up,\up}(q) = v^{\dn,\dn}(q), \\
    v^{\up,\dn}(q) = v^{\dn,\up}(q), \\
    v^{\up,\up}(q) \neq v^{\up,\dn}(q).
\end{align}
Defining the symmetric and antisymmetric interaction channels,
\begin{align}
    v^I(q)
    &= \frac{1}{2}
    \left[
        v^{\up,\up}(q)+v^{\up,\dn}(q)
    \right],
    \\
    v^z(q)
    &= \frac{1}{2}
    \left[
        v^{\up,\up}(q)-v^{\up,\dn}(q)
    \right],
\end{align}
the Hamiltonian of \cref{eq:projected-Hamiltonian-general} in terms of these variables takes the form
\begin{align}
    \bar H &= \bar H^I+\bar H^z,\\
    \bar H^I
    &=
    \frac{1}{2}
    \int\frac{\dd^2\bm q}{(2\pi)^2}\,
    v^I(q)
    \left[
        \bar\rho^I_{- q}\bar\rho^I_{ q}
        -
        N e^{-q^2/2}
    \right],
    \label{eq:H-I}\\
    \bar H^z
    &=
    \frac{1}{2}
    \int\frac{\dd^2\bm q}{(2\pi)^2}\,
    v^z(q)
    \left[
        \bar\rho^z_{- q}\bar\rho^z_{ q}
        -
        N e^{-q^2/2}
    \right].
    \label{eq:H-z}
\end{align}
Here,
    $\bar\rho^I_{ q}
    =
    \bar\rho^\up_{ q}+\bar\rho^\dn_{ q}$ is the total/symmetric density,
  and  
    $\bar\rho^z_{ q}
    =
    \bar\rho^\up_{ q}-\bar\rho^\dn_{ q}$ is the anti-symmetric density. The part $\bar H^I$ is $SU(2)$ symmetric, whereas $\bar H^z$ explicitly breaks the $SU(2)$ symmetry. The $SU(2)$ symmetric limit that we worked with in the majority of the main text corresponds to $v^{\up,\up}(q) = v^{\up,\dn}(q)$, i.e., $v^z(q)=0$, and in that setting [e.g., Coulomb interaction with real spins] we get back the $SU(2)$ symmetric gaps presented in the main text (see \cref{sec:collective} of the main text). For the bilayer FQH system, we consider the Coulomb interaction, $v^{\up,\up}(q) = 2\pi/q$ and $v^{\up,\dn}(q) = 2\pi e^{-q d}/q$, where $d$ is the interlayer separation.
    
\begin{figure*}[htpb]
    \centering
    \includegraphics[width=\linewidth]{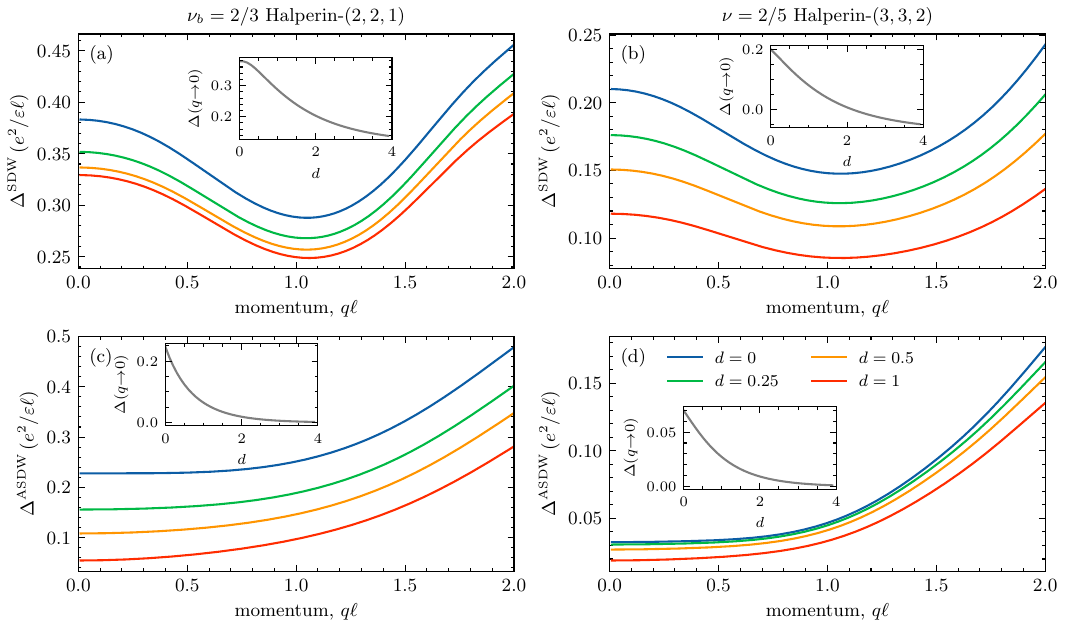}
  \caption{\textbf{Symmetric density wave (SDW) and anti-symmetric density wave (ASDW) gaps in bilayers.} Panels (a) and (c) show the SDW and ASDW gaps for the Halperin-$(2,2,1)$ state, while panels (b) and (d) show the corresponding gaps for the Halperin-$(3,3,2)$ state, calculated using \cref{eq:charge-oscillator-strength,eq:spin-oscillator-strength}. For the intra-layer Coulomb interactions, we have $v^{\up,\up}(q)=2\pi/q=v^{\dn,\dn}$, while for the inter-layer Coulomb interaction, we have $v^{\up,\dn}(q)=2\pi e^{-qd}/q=v^{\dn, \up}(q)$, where $d$ is the interlayer separation. The insets show the $q\to0$ behavior of the corresponding gaps as a function of $d$.    }
    \label{fig:SU2-broken-SDW-ASDW}
\end{figure*}
\subsection{Symmetric and anti-symmetric density-wave modes}
The density-wave gaps in the absence of $SU(2)$-symmetry of the Hamiltonian are computed following ideas similar to those presented in \cref{sec:collective} by using the commutator algebra of the appropriate density operators $\{\rh{\alpha}{q}\}$ [see \cref{eq:spinful-commutator-algebra}]. The resulting oscillator strengths for the $SU(2)$-broken interaction are given by 
\begin{align}
    \bar F(k)
    ={}&
    2\int\frac{\dd^2\bm q}{(2\pi)^2}\,
    v^I(q)\,
    \swedgesq{q}{k} \times \nonumber\\
    &\left[
        e^{\bm k\cdot\bm q}
        \bar S(|\bm k+\bm q|)
        -
        e^{-k^2/2}\bar S(q)
    \right]
    \nonumber\\
    &+
    2\int\frac{\dd^2\bm q}{(2\pi)^2}\,
    v^z(q)\,
    \swedgesq{q}{k} \times \nonumber\\
    &\left[
        e^{\bm k\cdot\bm q}
        \bar S^z(|\bm k+\bm q|)
        -
        e^{-k^2/2}\bar S^z(q)
    \right],
    \label{eq:charge-oscillator-strength} 
\end{align}
and
\begin{align}
        \bar F^z(q)
    ={}&
    2\int\frac{\dd^2\bm k}{(2\pi)^2}\,
    v^I(k)\,
    \swedgesq{k}{q} \times \nonumber \\
    &\left[
        e^{\bm q\cdot\bm k}
        \bar S^z(|\bm k+\bm q|)
        -
        e^{-q^2/2}\bar S(k)
    \right]
    \nonumber\\
    &+
    2\int\frac{\dd^2\bm k}{(2\pi)^2}\,
    v^z(k)\,
    \swedgesq{k}{q} \times \nonumber \\
    &\left[
        e^{\bm q\cdot\bm k}
        \bar S(|\bm k+\bm q|)
        -
        e^{-q^2/2}\bar S^z(k)
    \right].
    \label{eq:spin-oscillator-strength}
\end{align}
We show the symmetric and anti-symmetric density-wave gaps for the Halperin-$(2,2,1)$ and Halperin-$(3,3,2)$ states in a bilayer system in \cref{fig:SU2-broken-SDW-ASDW}. As anticipated, the gaps soften as the interlayer separation $d$ increases since with that the interlayer interaction is softened. The insets show the $q\to0$ gaps as a function of $d$. Since $2\pi/q - 2\pi e^{-qd}/q \approx 2\pi d$ at sufficiently small $q d$, using first order perturbation theory, the difference in the gaps for the $SU(2)$-symmetric Coulomb and $SU(2)$-breaking bilayer interaction scales linearly with $d$ and is independent of $q$ at $q d{\ll}1$. Similarly, just the gap for the $SU(2)$-breaking bilayer interaction goes down linearly with $d$ at sufficiently small $d$, consistent with the results shown in \cref{fig:SU2-broken-SDW-ASDW}.

\subsection{Spin-flip density wave mode}
\begin{figure*}[htpb]
    \centering
    \includegraphics[width=\linewidth]{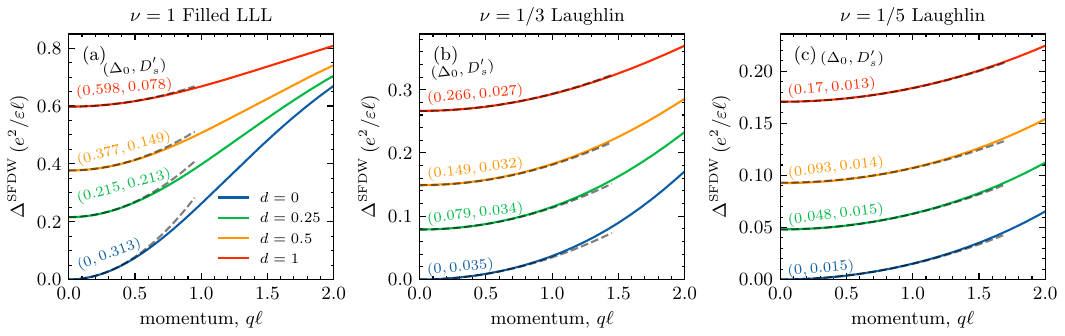}
    \includegraphics[width=\linewidth]{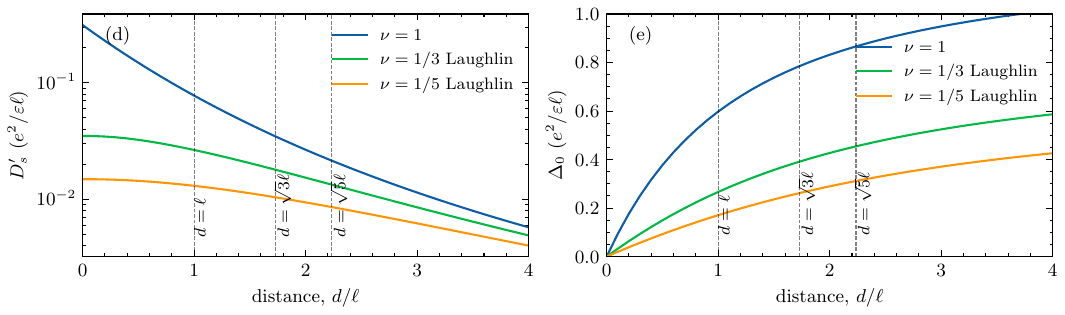}
\caption{\textbf{Spin-flip density wave (SFDW) gaps in bilayers.} 
Panels (a), (b), and (c) show the SFDW gaps for the $\nu=1$ integer quantum Hall, $1/3$ Laughlin, and $1/5$ Laughlin states, respectively, calculated using \cref{eq:SFDW-gap-SU2-broken}. For the intra-layer Coulomb interactions, we have $v^{\up,\up}(q)=2\pi/q=v^{\dn,\dn}$, while for the inter-layer Coulomb interaction, we have $v^{\up,\dn}(q)=2\pi e^{-qd}/q=v^{\dn, \up}(q)$, where $d$ is the interlayer separation. For $d\neq0$, the SFDW modes acquire a finite gap. The dashed lines show the quadratic fit $\Delta_0 + D_s' q^2$, obtained using \cref{eq: small_q_SFDW_v_1} for $\nu=1$, and \cref{eq: Ds_SU2_breaking,eq: SFDW_SU2_breaking_offset} for $\nu=1/3$ and $1/5$ Laughlin states. Panels (d) and (e) show $D_s'$ and $\Delta_0$, respectively, as a function of $d$.}
    \label{fig:SU2-broken-SFDW}
\end{figure*}

For a polarized state [by this we mean pseudospin/layer-polarized, since we assume the spins are always polarized], the spin-flip density wave excitation gap consequently is given by (up to the Zeeman term)
\begin{align}
    \Delta^{\mathrm{SFDW}}(q)
    ={}&
    \int_0^\infty
    \frac{\dd k}{2\pi}
    \,k\,
    v^I(k)
    \left[1-J_0(qk)\right]
    \left[1-S(k)\right]
    \nonumber\\
    &+
    \int_0^\infty
    \frac{\dd k}{2\pi}
    \,k\,
    v^z(k)
    \left[1+J_0(qk)\right]
    \left[1-S(k)\right].
    \label{eq:SFDW-gap-SU2-broken}
\end{align}
The first contribution is controlled by the $SU(2)$-symmetric interaction channel, while the second one arises solely from the symmetry-breaking component of the interaction. Taking $q\to0$, we see that 
\begin{align}
\label{eq: SFDW_SU2_breaking}
    &\Delta^{\mathrm{SFDW}}(q\to0) = \Delta_{0} + D^{'}_s q^2, \\
    \label{eq: SFDW_SU2_breaking_offset}
    &\Delta_{0} = 
    \int_0^\infty \frac{\dd k}{\pi}\,k\,
    v^z(k)
    \left[1-S(k)\right],\\
    \label{eq: Ds_SU2_breaking}
    &D^{'}_s
    =
    \int_0^{\infty}
    \frac{\dd k}{8\pi}
    \,k^3
    \left[v^I(k)-v^z(k)\right]
    \left[1-S(k)\right].
\end{align}
Here, $D^{'}_s$ is the modified spin-stiffness in the presence of the $SU(2)$-symmetry-breaking term. In \cref{fig:SU2-broken-SFDW}, we show the SFDW gaps for the $\nu=1$ filled LLL and the $\nu=1/3$ and $1/5$ Laughlin states for several values of $d$. As can be seen from Eq.~\eqref{eq: SFDW_SU2_breaking} [the appearance of a finite gap $\Delta_{0}$ that depends on $d$] and Fig.~\ref{fig:SU2-broken-SFDW}, the effect of the symmetry-breaking is to gap out the SFDW mode. This is anticipated, since in the symmetry-preserving case, the SFDW mode was gapless as it arose as a Goldstone mode out of the \emph{spontaneous} breaking of the continuous $SU(2)$ symmetry. Here, the interaction itself \emph{explicitly} breaks the $SU(2)$ symmetry. 

For $\nu=1$, using \cref{eq: SFDW_SU2_breaking_offset,eq: Ds_SU2_breaking} we get 
\begin{subequations}
\label{eq: small_q_SFDW_v_1}
\begin{align}
\label{eq: small_q_SFDW_v_1_Delta0}
    \Delta_{0}^{\nu=1} &= \sqrt{\frac{\pi }{2}} \left(1-e^{\frac{d^2}{2}} \text{erfc}\left(\frac{d}{\sqrt{2}}\right)\right),\\
\label{eq: small_q_SFDW_v_1_Dsp}
  [D_s']^{\nu=1} &=  \frac{1}{8} \left(\sqrt{2 \pi } \left(d^2+1\right) e^{\frac{d^2}{2}} \text{erfc}\left(\frac{d}{\sqrt{2}}\right)-2 d\right),
\end{align}
\end{subequations}
where $\text{erfc}(x)$ is the complementary error function, defined as $\text{erfc}(x){=}(2/\sqrt{\pi})\int_{x}^{\infty}e^{-p^2}dp$, and $d$ is the layer separation. Thus, in the long-wavelength limit, the SFDW gap at $\nu=1$ is [see \cref{eq: SFDW_SU2_breaking}]
\begin{eqnarray}
\label{eq: small_q_SFDW_v_1_gap}
    &&\Delta^{\mathrm{SFDW}}_{\nu=1}(q\to0) = \Delta_{0} + D^{'}_s q^2\\
    &=&\frac{1}{8} \left(\sqrt{2 \pi } e^{\frac{d^2}{2}} \text{erfc}\left(\frac{d}{\sqrt{2}}\right) \left(\left(d^2+1\right) q^2-4\right)-2 d q^2+4 \sqrt{2 \pi }\right) \nonumber.
\end{eqnarray}
At small-$d$, i.e., $d\to0$ or $d\ll \ell$, the $\nu=1$ SFDW gap is 
\begin{equation}
\label{eq: small_d_SFDW_v_1}
    \Delta^{\mathrm{SFDW}}_{\nu=1}(d\ll \ell) = d e^{-\frac{q^2}{2}}+\sqrt{\frac{\pi }{2}}  \left[1-e^{-\frac{q^2}{4}} I_0\left(\frac{q^2}{4}\right)\right].
\end{equation}

Taking the $q\to0$ limit of \cref{eq: small_d_SFDW_v_1}, in the small-$d$ limit at $\nu=1$, $\Delta_{0}^{\nu=1}(d\ll \ell)=d$, and the modified spin-stiffness is $[D^{'}_s]^{\nu=1}(d \ll \ell)=(\sqrt{2\pi}-4d)/8$ [Equivalently, one can obtain these by taking the $d\to0$ limit in \cref{eq: small_q_SFDW_v_1_Delta0,eq: small_q_SFDW_v_1_Dsp}.]. For $d=0$, the gap in Eq.~\eqref{eq: small_d_SFDW_v_1} reduces to the Kallin-Halperin $\nu=1$ spin-wave gap~\cite{Kallin84} quoted in Eq.~\eqref{eq: SFDW_nu_1}. 

The modified spin-stiffness $D^{'}_s$ is significantly reduced from its bare Coulomb value $D_s$ owing to the softening of the interaction due to the layer separation only when $d$ is comparable to the effective magnetic length sensed by the composite fermions $\ell^{*}=\sqrt{2p+1}\ell$~\cite{Jain07}. In other words, the spin-stiffness for the $\nu=1$ integer quantum Hall state changes readily with $d$, but for $1/3$ and $1/5$ Laughlin states, one needs to go to larger $d$ that is comparable to $\sqrt{3}\ell$ and $\sqrt{5}\ell$, respectively, before the spin-stiffness decreases significantly from its Coulomb value [see Fig.~\ref{fig:SU2-broken-SFDW}].

\bibliography{biblio_fqhe1}

\end{document}